\documentclass[conference]{IEEEtran}

\usepackage{graphicx}
\usepackage{subcaption}
\usepackage{booktabs}
\usepackage{color}
\usepackage[ruled,vlined,linesnumbered]{algorithm2e}
\usepackage{tikz}

\begin{document}
\title{SIMD Lossy Compression for Scientific Data}

\author{\IEEEauthorblockN{Griffin Dube\IEEEauthorrefmark{1},
Jiannan Tian\IEEEauthorrefmark{2},
Sheng Di\IEEEauthorrefmark{3},
Dingwen Tao\IEEEauthorrefmark{2},
Jon C. Calhoun\IEEEauthorrefmark{1}, 
Franck Cappello\IEEEauthorrefmark{3}}
\IEEEauthorblockA{\IEEEauthorrefmark{1}Holcombe Department of Electrical and Computing Engineering -- Clemson University, Clemson, SC 29634}
\IEEEauthorblockA{\IEEEauthorrefmark{2}Department of Computer Science -- Washington State University, Pullman, WA 99164}
\IEEEauthorblockA{\IEEEauthorrefmark{3}Argonne National Laboratory, Lemont, IL 60439}}
\maketitle

\begin{abstract}
Modern HPC applications produce increasingly large amounts of data, which limits the performance of current extreme-scale systems. Data reduction techniques, such as lossy compression, help to mitigate this issue by decreasing the size of data generated by these applications. SZ, a current state-of-the-art lossy compressor, is able to achieve high compression ratios, but the prediction/quantization methods used introduce dependencies which prevent parallelizing this step of the compression. Recent work proposes a parallel dual prediction/quantization algorithm for GPUs which removes these dependencies. However, some HPC systems and applications do not use GPUs, and could still benefit from the fine-grained parallelism of this method. Using the dual-quantization technique, we implement and optimize a SIMD vectorized CPU version of SZ, and create a heuristic for selecting the optimal block size and vector length. We also investigate the effect of non-zero block padding values to decrease the number of unpredictable values along compression block borders. We measure performance of vecSZ against an O3 optimized CPU version of SZ using dual-quantization, pSZ, as well as SZ-1.4. We evaluate our vectorized version, vecSZ, on the Intel Skylake and AMD Rome architectures using real-world scientific datasets. We find that applying alternative padding reduces the number of outliers by 100\% for some configurations.
Our implementation also results in up to 32\% improvement in rate-distortion and up to 15$\times$ speedup over SZ-1.4, achieving a prediction and quantization bandwidth in excess of 3.4 GB/s.
\end{abstract}



\maketitle

\section{Introduction} \label{sec:intro}
As data produced by large scale scientific applications becomes larger, efficient management of application data in high-performance computing (HPC) systems is becoming increasingly important. Current petascale applications such as the Hardware/Hybrid Accelerated Cosmology Code (HACC)~\cite{HACC} can produce 21.2 petabytes of data when simulating 2 trillion particles for 500 time-steps. These large amounts of data are difficult or impossible to handle due to the limitations of I/O bandwidth and storage on modern HPC systems. One mechanism for coping with the massive amounts of data generated by these applications is through the use of data reduction techniques such as data compression. 

Data compression is broken down into two areas: lossless and lossy. Lossless compression reduces data size and preserves the original data exactly after decompression. However, it is only able to achieve limited compression ratios 1--4$\times$ on floating-point HPC datasets~\cite{compressionSurvey2014}.
Lossy compression is able to achieve higher compression ratios than lossless compression by introducing error into data~\cite{SZ, ZFP}. Error-bounded lossy compression (EBLC) presents an attractive solution to the data reduction challenge because of its ability to achieve high compression ratios while guaranteeing the error introduced remains within a specified error bound. 
The ability to tune the level of loss in the data enables lossy compression to be integrated into HPC applications and workflows making them more efficient~\cite{frank_survey, CalhounIJHPCA:LossyCompression, szQuantumSim, LINDSTROM_seismic}.

SZ is a popular EBLC algorithm seeing rapid development lately.
SZ supports a variety of error bounding modes~---~e.g., absolute and relative error~\cite{SZ}, PSNR~\cite{SZ_PSNR}. Newer versions of SZ optimize the compression ratio and the compression/decompression bandwidth~\cite{Pavlo:Cluster2019:lossyCR}; however, to significantly improve the compression/decompression bandwidth, SZ must take advantage of accelerators~\cite{GhostSZ, cusz}. The CPU version of SZ is limited to coarse grain parallelism for its prediction and quantization process. This is due to a read-after-write (RAW) dependency that prevents fine grained parallelism necessary to perform SIMD operations. A GPU implementation of SZ, cuSZ~\cite{cusz}, addresses this by introducing a dual-quantization, or dual-quant, method that removes the RAW dependency and allows for fine grained parallelism in this step of compression. However, dual-quant has not been explored for CPUs resulting in unrealized performance.

In this paper, we investigate the performance of cuSZ's dual-quant technique when applied to CPUs while optimizing our own version of SZ, that uses the dual-quant method for prediction and quantization of data values. 
We use this implementation as a basis for investigation of further optimizations through vectorization, autotuning and modification of block padding to improve both compression bandwidth as well as compression ratio. Our new version of SZ, called vecSZ, leverages fine-grained parallelism on the CPU enabling significantly better compression/decompression bandwidths. 

In this paper, we:
\begin{itemize}
    \item Generate a Roofline performance model to demonstrate the peak CPU performance of the dual-quant algorithm, showing that a sequential implementation of the algorithm only reaches up to 25\% of peak on CPU architectures;
    
    \item Apply vectorization and threading to the dual-quant prediction algorithm for SZ increasing the prediction/quantization bandwidth by 15.1$\times$;
    
    \item Develop an auto-tuning framework to select the vector length and compression block size to achieve near optimal performance across multiple time-steps and error bounds; and
    
    
    \item Investigate the use of non-zero padding values for compression block borders, reducing the overall number of unpredictable values by as much as 100\%, leading to a 32\% improvement in rate-distortion.

\end{itemize}


\section{Background and Related Work} \label{sec:background}

\subsection{Lossy Data Reduction}

Data sets in HPC contain large amounts of floating-point data. Due to the seemingly random nature of mantissa bits, standard lossless compression methods do not give a significant level of reduction. Large levels of reduction come at the expense of data fidelity. Instead of saving data at each time-step,  decimation stores data from a subset of time steps, often in full resolution. Truncation is a naive form lossy compression that lowers the precision level of the data~---~e.g., 64-bit to 32-bit floating-point. Data sampling reduces the number of data values saved in a single data array by sampling them at random~\cite{Woodring_2011, 
wei2018information} or based on some complex criteria~\cite{
databig2,Biswas_2018}, but requires reconstruction algorithms to expand the samples to the full resolution~\cite{1163711}.
To control the accuracy in the reduced data error-bounded lossy compression algorithms such as SZ~\cite{SZ} and ZFP~\cite{ZFP} provide order-of-magnitude larger reductions than lossless compression while meeting a users specified level of data fidelity. However, setting the compressor's error bound to ensure data fidelity is an open question~\cite{CalhounIJHPCA:LossyCompression, Baker2014:HPDC,Foresight}.


\subsection{SZ}
SZ \cite{SZ, SZ_PSNR} is an error-bounded lossy compressor that compresses a data set by first decomposing the data into fixed sized blocks. SZ then compresses each block in a multi-step process: 
(1) Data Prediction -- the data points in each block are predicted based on previously predicted values using either a Lorenzo predictor or a local linear regression predictor; 
(2) linear-scale quantization --  the error in the predicted value for each data point is converted from a floating-point value to an integer by applying a equal-bin-size quantization between the range of $[-\epsilon, \epsilon]$, where $\epsilon$ is the compressor's error bound; and
(3) encoding -- the sequence of integer codes are further compressed using entropy encoding techniques such as Huffman coding or dictionary based methods such as GZip~\cite{gzip} and Zstd~\cite{zstd}.
SZ bounds the induced error in multiple ways: absolute error bound or value-range relative error bound 
\cite{SZ}, 
target PSNR 
\cite{SZ_PSNR}, etc.
%
%
SZ supports multiple I/O libraries such as HDF5, NetCDF, and Adios2. In addition, SZ takes advantage of on-node parallelization such as OpenMP, GPUs~\cite{cusz}.


\subsection{SIMD}

Many operations in HPC applications such as linear algebra exhibit large amounts of data parallelism. The addition of two vectors $a$ and $b$ of length $n$ is decomposed into  $n$ independent addition operations that can be done in parallel. This type of computation efficiently maps to single instruction multiple data (SIMD)
Modern HPC systems leverage SIMD parallelism on GPUs to achieve high FLOP rates. The current generation of CPUs incorporate vector extensions such as Streaming SIMD Extensions, Advanced Vector Extensions (AVX), and on recent Intel CPUs AVX-512 to accelerate data parallel computation.

\section{Performance Optimization} \label{sec:methods}
Current CPU implementations of SZ are designed to give large compression ratios at reasonable compression bandwidths. To improve the compression bandwidth, SZ employs thread-level parallelization via OpenMP where each thread works independently on a number of blocks. The current CPU algorithm has a read after write (RAW) dependence which precludes optimization via fine-grained parallelism. Failure to take advantage of fine-grained parallelism limits SZ's compression bandwidth. We enable fine-grained parallelism by exploiting the dual-quant algorithm of cuSZ~\cite{cusz} for data prediction and error quantization. By using the dual-quant algorithm we are able to apply SIMD parallelism to CPU version of SZ during the data prediction and error quantization step. We focus on this step as opposed to the Huffman encoding step because there exist vectorized implementations of Huffman encoding \cite{huffman_code_SIMD} whereas the dual-quant operation has been used for GPUs in \cite{cusz}, and not CPU vectorization.


\subsection{Dual-Quantization}\label{sec:dual-quant}

SZ lossy compression chunks the original data set $D$ into fixed sized blocks. Each decomposed block is able to be compressed independently. Algorithm~\ref{algo:original_sz} shows an overview of the compression and decompression processes. We denote variables we generate and use during compression with an open circle superscript. Furthermore, for data in decompression we denote them with a closed circle superscript. 

Compression for each data point $d$ in the full data set $D$ begins by predicting the data value via Lorenzo prediction $\ell$. Lorenzo prediction predicts the value of $d$ based on the values of previously predicted surrounding data, $d_\textsc{\scshape sr}$, in the block\footnote{Details of the Lorenzo predictor are found in~\cite{lorenzo,SZ_Paper2}.}. After prediction, we compute the error 
$e^\circ$ in the original data $d$ and our prediction
$p^\circ$. Next, we quantize the error based on the user-selected error bound $eb$. Quantizing the error allows us to represent the error in the prediction as an integer, which compresses more efficiently than a floating-point value. Data whose prediction error is larger than the error bound is stored in the \texttt{else} block verbatim with no loss in accuracy. Decompression uses the Lorenzo prediction to reconstruct the, reversing of the quantization process.

\begin{figure}[h]
    \centering
    \includegraphics[width=\columnwidth]{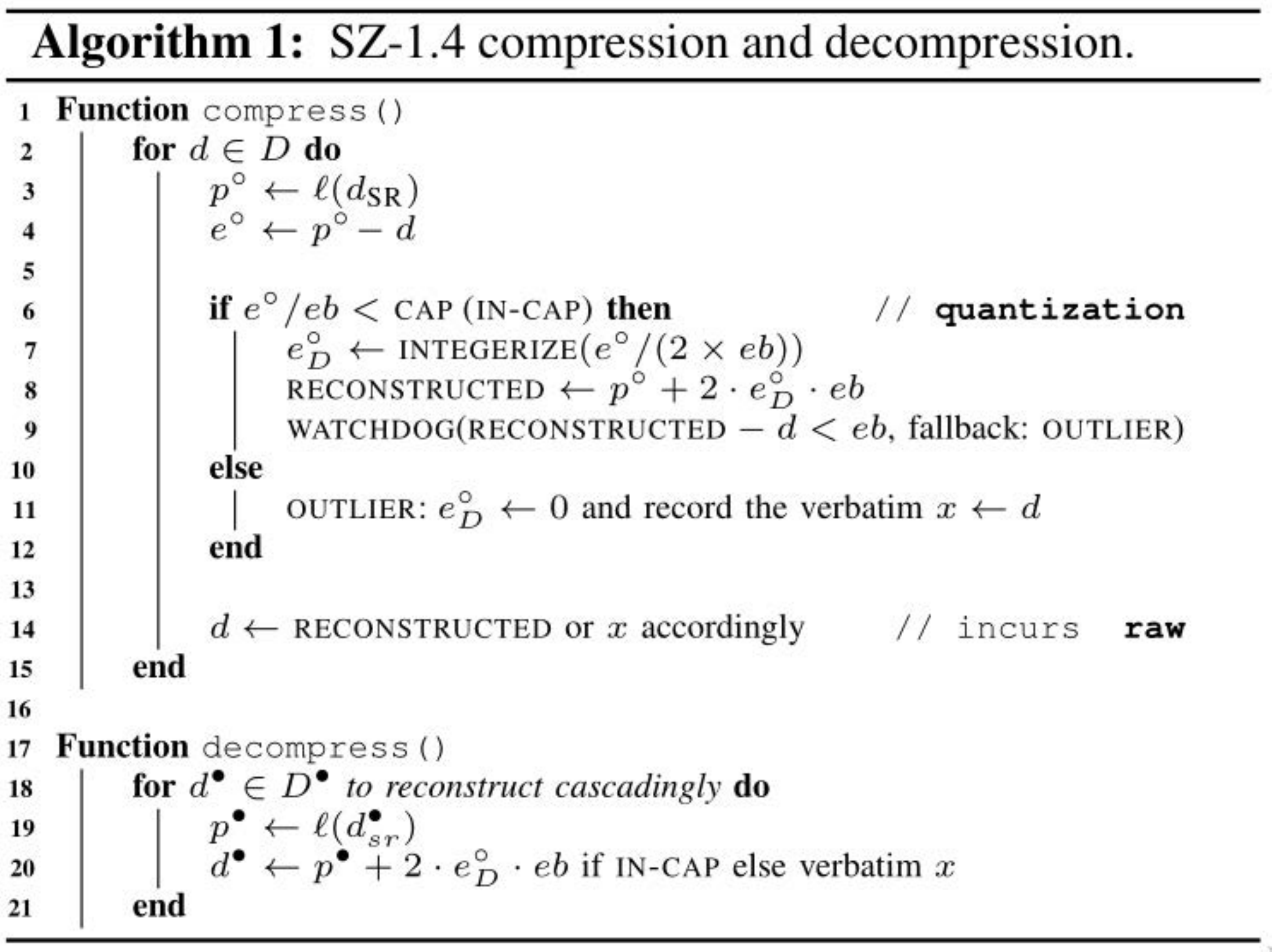}
\end{figure}

\newcounter{algo}[section]
\refstepcounter{algo}\label{algo:original_sz}
\refstepcounter{algo}\label{algo:dual-quant}
As written, Algorithm~\ref{algo:original_sz} has a loop carried RAW dependence on line 14. Because at the next iteration, the Lorenzo predictor needs predicted data values from prior iterations, the SZ-1.4 algorithm is not able to be parallelized by vectorization. To remove the RAW dependence, the \textit{dual-quant} algorithm (Algorithm~\ref{algo:dual-quant}) separates the prediction and quantization step, allowing for fine-grained parallelization~\cite{cusz}.

Lines 2 and 3 represent the pre-quantization stage, where each datum $d \in D$ is quantized based on the user's selected error bound $eb$ forming a new dataset $D^\circ$ of each quantized datum $d^\circ$. This new floating-point dataset $D^\circ$ has an error that is less than the user's error bound $|d - 2\cdot d^\circ \cdot eb| < eb$. After the pre-quantization step, we use Lorenzo prediction to predict the values of $d^\circ$. 

After the pre-quantization step completes, the dual-quant algorithm starts the post-quantization step. In this step, we compute the difference, $\delta^\circ$, between the predicted value and the prequantized value, $d^\circ$. We quantize $d^\circ$ similarly to linear-quantization in SZ-1.4 to obtain an integer quantization code. Because we have pre-computed all of the prediction values, we are able to quantize all the predictions in parallel removing the RAW dependence for the compression dual-quant operation. In this paper, we focus on vectorization of dual-quant for compression and not decompression, because decompression still carries a dependency in that each data point cannot be decompressed until the values preceding it are reconstructed. 

\begin{figure}[h]
    \centering
    \includegraphics[width=\columnwidth]{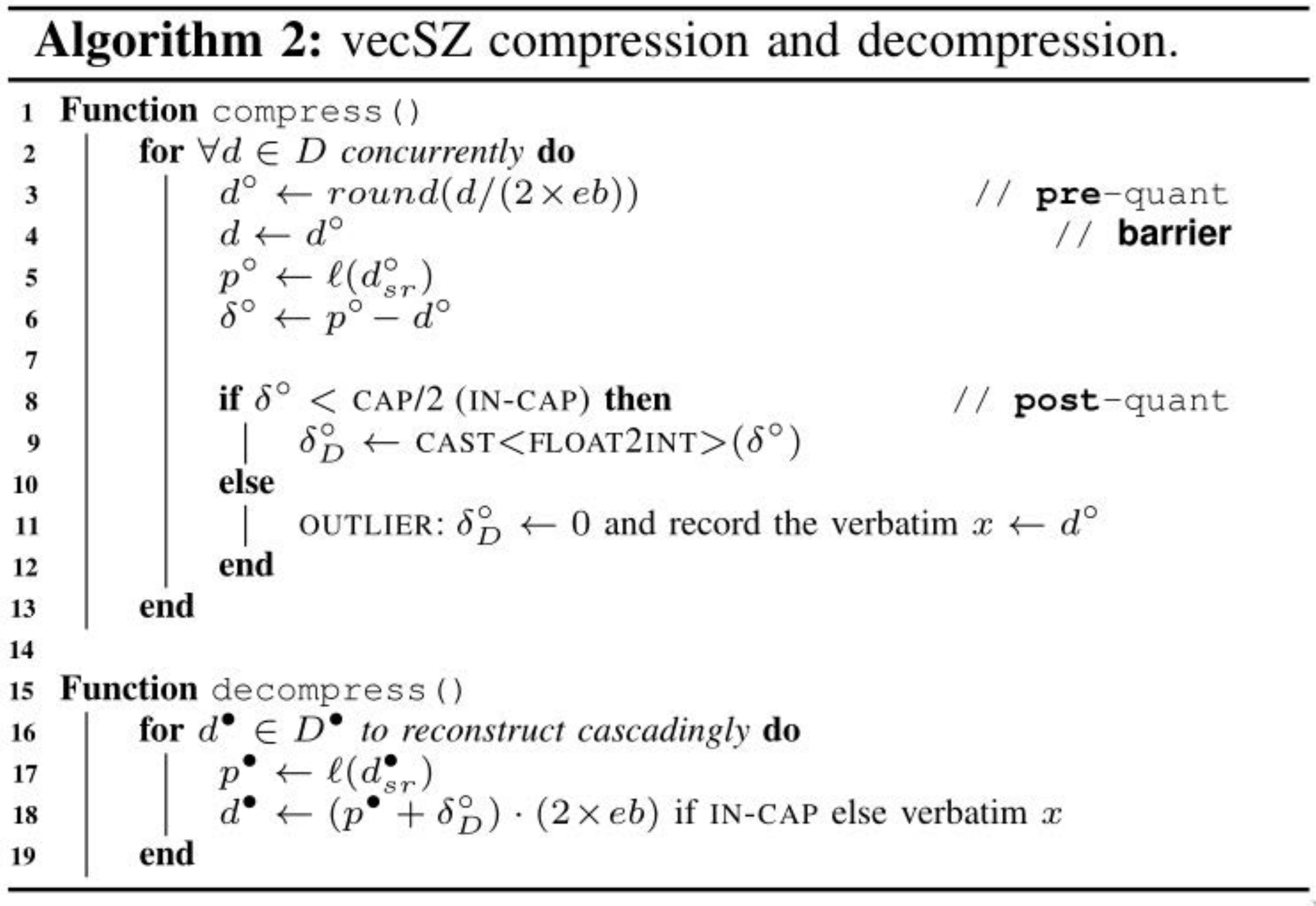}
\end{figure}

\subsection{Performance Modeling}\label{sec:roofline}

When optimizing a code, programmers seek the highest level of performance. However, the maximal rate of computation for a piece of hardware is often an unrealistic goal. The true performance of an algorithm and its ability to reach the maximal computation rate is highly dependent on if the algorithm is computation or memory bound.

To establish the maximal performance level, we construct a Roofline performance model for our hardware. The Roofline model is a way to visualize the peak floating-point performance and memory performance based on the operational intensity, or operations per byte of DRAM traffic, of a given algorithm \cite{rooflinemodel}.
In order to generate the ceilings for the roofine model, we find the sustainable memory bandwidth and the peak floating-point performance of the CPU. We use Lawrence Berkeley National Lab's Empirical Roofline Tool (ERT) \cite{ert} to determine these characteristics for each CPU we use. The ERT determines the machine characteristics by running micro-kernels on the target machine and generates the bandwidth and GFLOP/s data needed by the Roofline model. 

\begin{figure}[ht]
    \centering
    \includegraphics[width=\columnwidth, trim={90 16 35 16 }, clip]{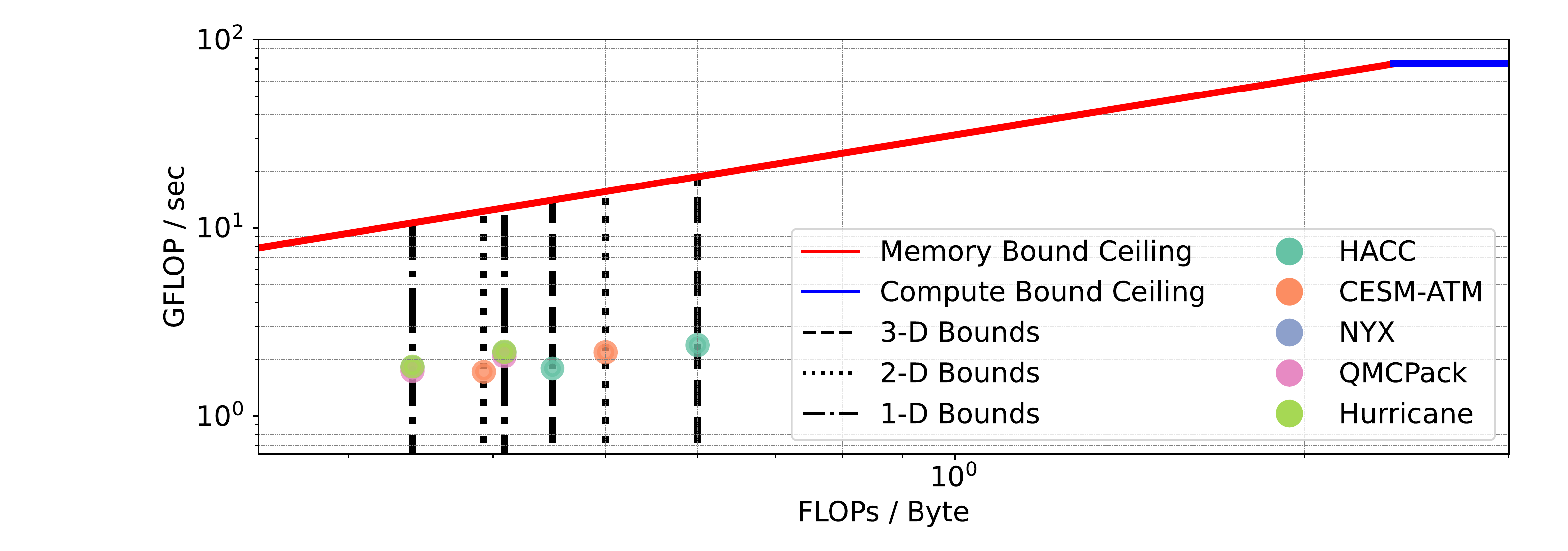}
    \caption{Roofline model of the operational intensities for pSZ 1D, 2D, and 3D dual-quant.}
    \label{fig:roofline_original}
\end{figure}

Analyzing the dual-quant code in Algorithm~\ref{algo:dual-quant}, we compute operational intensity (OI) as the number of floating point operations performed per byte accessed by DRAM and FLOPS as the number of floating point operations per second.
We generate a conservative and lenient bound on the OI of the dual-quant operation. The conservative bound is calculated by including strictly arithmetic operations when calculating the FLOPS/byte while the lenient bound also includes operations such as type casts and comparisons performed on floating point values. Choosing conservative and lenient bounds as opposed to a single OI ensures our actual performance lies between these two bounds. We find that for both the lenient and conservative estimates the OI for the dual-quant operation is memory-bound, corresponding to values under the slanted region of the model. Figure~\ref{fig:roofline_original} shows the conservative and lenient bounds for 1D, 2D, and 3D version of the dual-quant algorithm. We also direct attention to the performance of a sequential dual-quant implementation shown in this figure. Without any form of optimization, the dual-quant algorithm is unable to take advantage of the full amount of resources available on the CPU and reaches between 10-25\% of the theoretical peak performance of the machine. In the remainder of this section, we discuss the optimization techniques we use to vectorize and optimize the SZ prediction and quantization steps based on the dual-qual algorithm that removes the RAW dependency.

\subsection{Vectorization}
After implementing an algorithm without the loop-carried RAW dependency of the original SZ, we look at vectorization as a means of taking advantage of SIMD parallelism on the CPU. Using compiler based autovectorization presents a non-labor intensive way to apply vectorization without the need for further modification to the actual code, however autovectorization relies on certain properties of the code to be determined at compile time with respect to control flow that are not guaranteed by the dual-quant algorithm~\cite{Nuzman2006}. One example of such behavior is due to the manner in which we ensure that computation is performed only within the dimensions of our data, and any out-of-bounds computation as a result of partitioning data into blocks is discarded. In a sequential program, it is sufficient to continue to the next row or block if the remaining values in a block contain out-of-bounds elements. The control logic required to perform this operation prevents autovectorization from occurring. In a vectorized program, as long as the vector register contains at least one value that lies within the dimensions of our data, there is no additional cost to compute on these out-of-bounds elements. Thus, we modify the boundary check to perform fewer checks, at a vector register granularity as opposed to a data element granularity by generalizing \texttt{if} statements that check boundaries and moving them outside \texttt{for} loops. 

For the sections of code not autovectorized, we manually vectorize via GCC compiler intrinsics.
Using these intrinsic functions, we are also able to port this code to other CPUs that use AVX vectorization without modification to the code. Using the GCC intrinsic vector functions, we vectorize the pre- and post-quantization loops manually to perform work on up to 16 floating point values at a time for AVX-512 capable CPUs and 8 floating point values for AVX2 capable CPUs.

The manual vectorization of a portable dual-quant implementation capable of leveraging different levels of vector registers (AVX-512, AVX2, etc.) introduces several challenges that we must be consider. The AVX-512 intrinsics contain a wider range of instructions, so determining how to map these instructions to the operations available on CPUs with lower vector capabilities is important for ensuring performance does not degrade. In particular, we look at the latency and throughput of the instructions, selecting the most comparable one available for each of the sets of intrinsics~\cite{intrinsics}.

\subsection{Block Size}
Before SZ compresses data, it logically decomposes the data set into small fixed sized blocks which are compressed independently. The dimension of the block sizes is not configurable in the original SZ. However, when mapping the vector operations to the computation in each block and with certain block sizes, not all of the vector register is utilized, leading to inefficiencies. For example, a block size of $6 \times 6$ (2D data) and a vector width capable of holding 8 values leads to 25\% of the register not utilized for each operation. To reduce this inefficiency, we use block sizes that are multiples of the vector register in use. The traditional SZ block sizes of 256 for 1D, 16$\times$16 for 2D, and 6$\times$6$\times$6 for 3D leave additional space for more computation to be performed in vector registers of 256-bits and 512-bits in length. The optimal block size used by 1D, 2D, and 3D data varies based on the input provided, and the size of the vector registers available on the CPU, but we concentrate on block sizes of 8, 16, 32, and 64. We do not provide results for block sizes of 128 and 256 for vecSZ in order to improve the clarity of plots because we saw no improvement in performance.  

\subsection{Autotuning}
The performance between different configurations of block size and vector length has the potential to vary prediction and quantization bandwidth by up to 300\% depending on the dataset. To determine the optimal configuration of block size and vector register length to use for a particular input, we develop a heuristic for tuning these parameters by performing computation on a sample of random blocks using each of the configurations. Before running the dual-quant algorithm on all of the data, we perform an exhaustive search of all configurations, sampling a fixed percentage of blocks from the data set in order to estimate the optimal configuration of block size and vector length. We repeat this test multiple times and choose the best performing configuration to use during compression. We find that we can amortize the overhead of autotuning when running multiple time-steps of a simulation because the best configuration of block size and vector register length holds across the majority of simulation time-steps.

\subsection{OpenMP}
We introduce thread-level parallelism at the granularity of a single block using OpenMP version 4.5. Each block is calculated independently from all other blocks, so we introduce coarse-grained parallelism at the block level to further accelerate the fine-grained parallelism we apply through vectorization. We optimize the scheduling of threads on cores using OpenMP thread affinity controls. We use the environment variables \texttt{OMP\_PLACES=cores} and \texttt{OMP\_PROC\_BIND=close} to schedule threads to all cores on a single socket before beginning to schedule threads on the next socket \cite{omp_affinity}. This configuration ensures we keep threads as close to the data on which they are operating for as long as possible. While SZ currently only supports OpenMP for 3D data, we implement OpenMP capabilities in vecSZ for 1D, 2D, and 3D data.


\section{Prediction Optimization}
\begin{figure}[h]
    \centering
    \includegraphics[width=\linewidth]{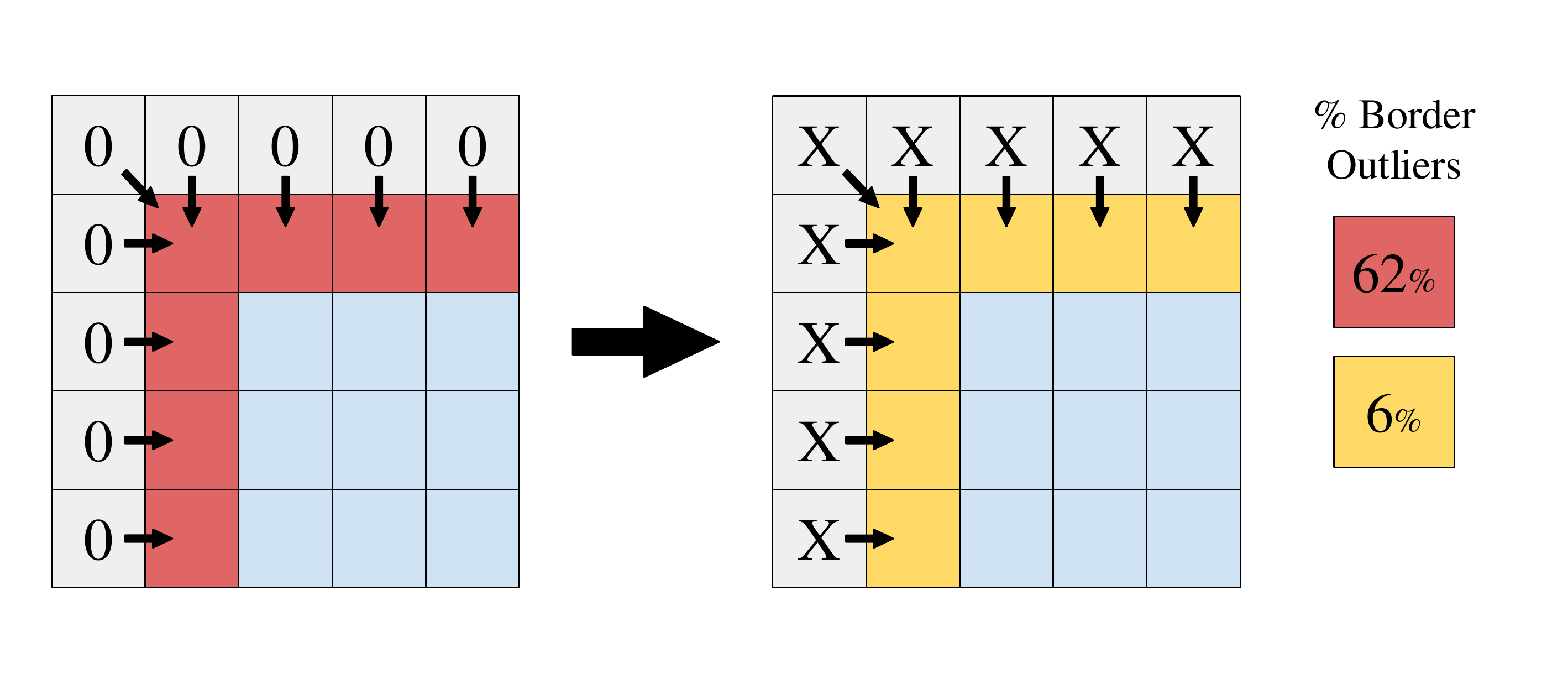}
    \caption{Example prediction of border values using zero-padding versus alternative padding. Sample values from CESM data, applying a block-based average can reduce the percentage of unpredictable data along borders by 91\%. }
    \label{fig:padding}
\end{figure}

We investigate the effect of values used in padding compression blocks on the accuracy of the dual-quant method.
In previous work \cite{cusz}, the values used for padding compression blocks are chosen to be all zeroes, regardless of the dataset being compressed. During the prediction and quantization stage of compression, values without preceding elements (i.e. those found along borders) rely on the block padding for prediction, as shown using data from a sample run of CESM's CLDHGH dataset as a 2-D example in Figure~\ref{fig:padding}. In extreme cases, 100\% of the unpredictable data points may lie on the border of the compression block.

Choosing a value such as zero provides inconsistent results across different datasets. Applying this technique to a near zero dataset results in more reliable prediction of data values along borders. On the other hand, a dataset with relatively few near zero values, like that shown in Figure~\ref{fig:padding}, could suffer from a large number of unpredictable data, or outliers, along block borders. Increasing the number of outliers, decreases the compression ratio because unpredictable data needs more storage per element than predictable data. 

In order to address this inconsistency, vecSZ chooses a padding value based on statistical properties of the data. This reduces the number of outliers found along block borders by selecting a padding value that has potential to more closely represent the data found at the borders of a compression block. Our methods of alternative padding can be applied to other prediction-based compression such as cuSZ as well.

\subsection{Determining Padding Value}
When computing padding values, we investigate the effect of choosing a minimum, maximum, or average value of our data as opposed to the traditional zero padding. We choose padding based on these properties because the borders of a given block are more accurately described by a statistical representation of the data than a constant value chosen at design time without knowledge of the data being compressed. 
It is possible to extend this implementation to include additional, more robust, methods for selecting a scalar value for padding; however, we find that our chosen operations are sufficient for demonstrating the benefit of alternative padding. 

\subsection{Padding Granularity}
\subsubsection{Global}
This case has the lowest overhead, due to the fact that only one additional data point must be stored in our compressed data to allow us to reconstruct the data. The sacrifice we make to achieve this low overhead is in the accuracy of the padding scalar chosen. Computing a single, global scalar does not provide the same level of control as finer granularity, and a global average is not necessarily the best representation of all data in the dataset.

\subsubsection{Block}
One scalar is computer per compression block, meaning we must store additional data points equivalent to the number of blocks in our compressed data. This results in a higher storage overhead, but provides finer control over the padding value chosen per block. This granularity is best suited for datasets that vary significantly across their domain. 

\subsubsection{Edge}
Edge granularity computes a separate scalar for each set of border values, resulting in $nBlocks * nDim$ additional data points. 
Although this provides the most fine-tuned control over padding values, in our experiments we find the overhead of an edge based scalar to outweigh it's benefit due to the number of additional data points needed to reconstruct data computed using this granularity of padding. 



\section{Experimental Results}\label{sec:results}

\subsection{Hardware}\label{sec:hardware}
We conduct our experiments using hardware available on CloudLab, a facility for building clouds that provides access to computing, storage, and networking resources \cite{cloudlab}. We concentrate our attention on two popular CPU models commonly found in HPC systems: the AMD EPYC Rome and the Intel Xeon Gold, described in detail in Table \ref{tab:hardware}. The Intel nodes we use contain two Intel Xeon Gold 6142 CPUs, and the AMD nodes contains a single AMD EPYC Rome 7452.
\begin{table}[ht]
    \centering
    \begin{tabular}{ccc}
        & \multicolumn{2}{c}{\textbf{CPU}}\\
        & \textit{AMD EPYC Rome 7452} & \textit{Intel Xeon Gold 6142} \\
        \toprule
        \multicolumn{1}{c}{\textbf{RAM}} & 127 GB & 384 GB\\
        \multicolumn{1}{c}{\textbf{Cache}} & 128 MB & 22 MB\\
        \multicolumn{1}{c}{\textbf{Cores (Threads)}} & 32 (64 SMT) & 16 (32 HT)\\
        \multicolumn{1}{c}{\textbf{Vectorization}} & AVX2 & AVX-512\\
        \bottomrule
    \end{tabular}
    \caption{Detailed hardware used for experiments (HT is Hyper-threading and SMT is Simultaneous Multi-Threading)}
    \label{tab:hardware}
\end{table}

A notable difference between the two CPU architectures is the level of vectorization supported on each CPU. The Intel Xeon Gold 6142 has AVX-512 support, meaning support for 512-bit vector registers capable of operating on 16 single-precision floating point values at a time, while the AMD EPYC Rome 7452 only supports operations on up to 256-bit vector registers or 8 single-precision floating point values at a time.

The cache size of the Intel CPU is significantly smaller than that of the AMD CPU. Although the Intel CPU is able to perform computation on double the number of values simultaneously, its smaller cache size means there is a higher probability that our code will be required to more cache misses on the Intel Gold CPU than on the AMD Rome CPU.

\subsection{Test Data Sets}
We test our vectorized code on a set of real world HPC data sets from the Scientific Data Reduction Benchmarks \cite{sdrbench}. The data sets we list in Table~\ref{tab:datasets} are representative of a wide range of HPC workloads of varying size, problem domain, and dimension. For the CESM-ATM data set, we use an absolute error bound of  $1{e-}5$. For all the remaining data sets, we use an absolute error bound of $1{e-}4$.

\begin{table}[ht]
    \centering
    \begin{tabular}{ccccc}
        \textbf{Data Set} & \textbf{Domain} & \textbf{Type} & \textbf{Dimensions} & \textbf{Size (MB)} \\ \toprule
        HACC & Cosmology & fp32 & 280,953,867 & 1071.75\\
        CESM & Climate & fp32 & 1,800$\times$3,600 & 24.72\\
        Hurricane & Climate & fp32 & 100$\times$500$\times$500 & 95.37\\
        NYX & Cosmology & fp32 & 512$\times$512$\times$512 & 512.00\\
        QMCPACK & Quantum & fp32 & 288$\times$115$\times$69$\times$69  & 601.52\\
        \bottomrule
    \end{tabular}

    \caption{Attributes of the data sets used in experiments. HACC (6 fields), CESM (3 fields), Hurricane (20 fields), NYX (6 fields), and QMCPACK (2 fields).}
    \label{tab:datasets}
\end{table}

\subsection{Experimental Methodology}
We use the C++ \texttt{ctime} library's high resolution clock for timing the compression, dual-quant, and autotuning operations as well as the total runtime of our code. Each experiment is run on both AMD and Intel CPUs and results are averaged across ten total runs of each dataset. We plot the standard deviation as error bars when appropriate. As a baseline for evaluating vecSZ we use pSZ, a serial version of SZ that uses the dual-quant method as opposed to the prediction and quantization method used by SZ-1.4.13.5  \cite{SZ_Paper2}. We compare performance of vecSZ to the performance of SZ-1.4 as opposed to SZ-2.1 because the prediction and quantization method used in SZ-1.4 is directly comparable to the dual-quant operation applied in vecSZ as it always uses Lorenzo prediction whereas the linear regression methods used in SZ-2.1 do not provide a fair performance comparison. For each version of SZ, we use the same standard config file that comes with SZ-1.4.13.5.  GCC 9.3.1's \texttt{-O3} option compiles and optimizes all of our codes with OpenMP version 4.5

\subsection{Comparison to SZ}
 
\begin{figure}[h]
    \begin{subfigure}{.49\linewidth}
        \centering
        \includegraphics[width=\linewidth]{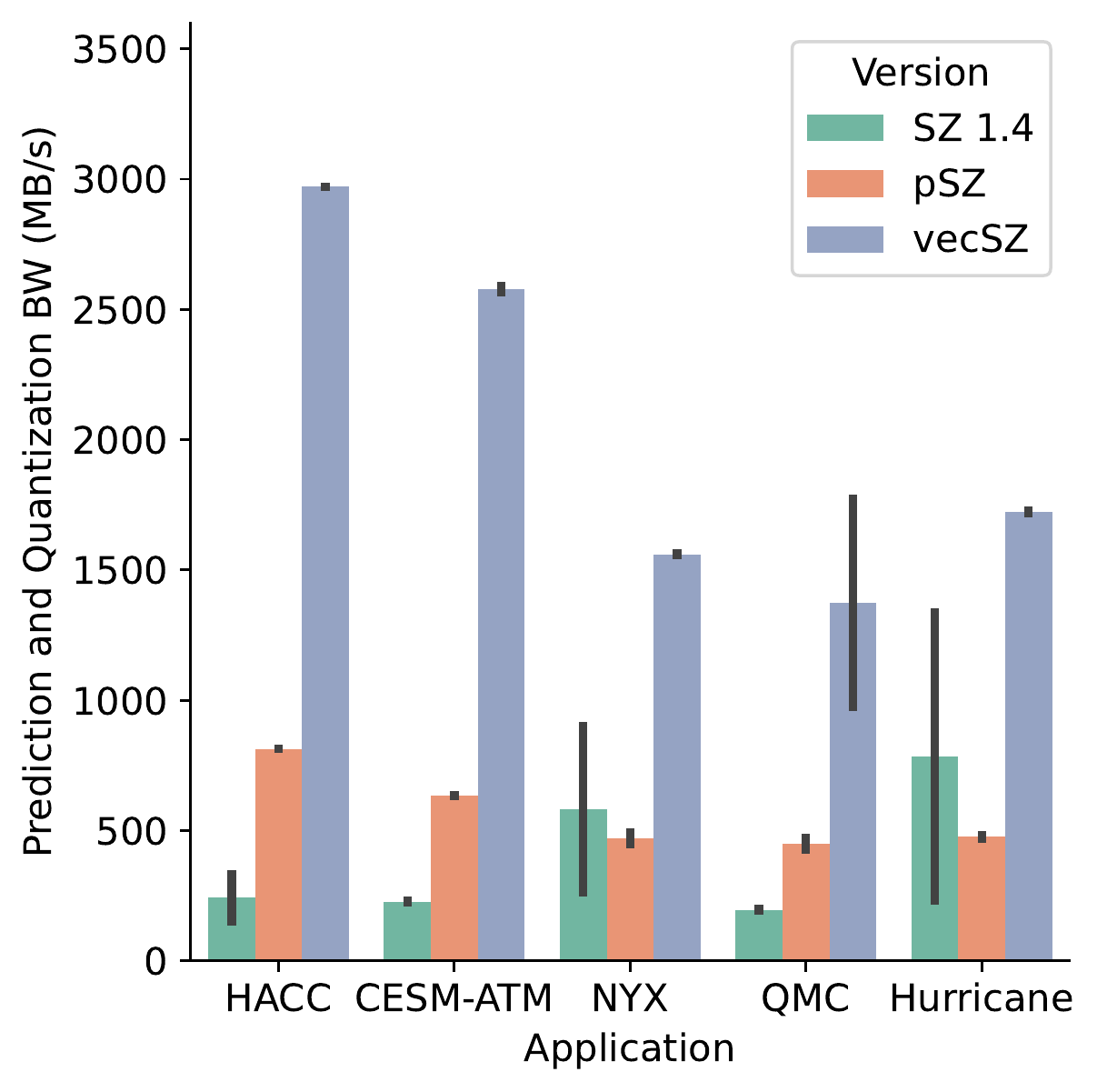}
        \caption{AMD}
        \label{fig:version_comparison_amd}
    \end{subfigure}
    \begin{subfigure}{.49\linewidth}
        \centering
        \includegraphics[width=\linewidth]{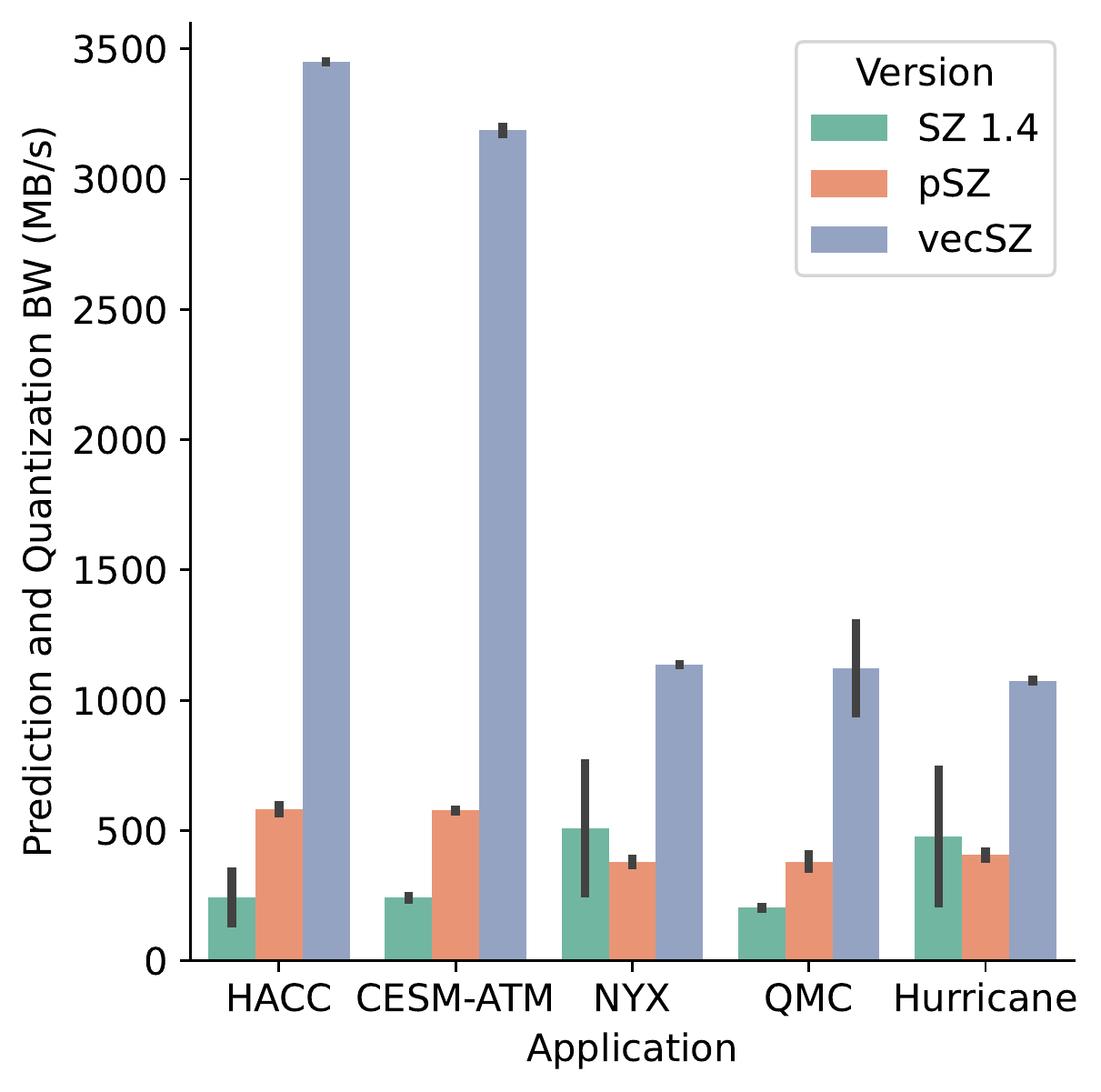}
        \caption{Intel}
        \label{fig:version_comparison_intel}
    \end{subfigure}
        \caption{Prediction and quantization bandwidth of SZ-1.4, pSZ, and vecSZ. Black bars are standard deviation.}
        \label{fig:version_comparison}
\end{figure}

Applying vectorization to the dual-quant method enables it to process data faster, leading to improvements in bandwidth. Figure~\ref{fig:version_comparison} shows the prediction and quantization bandwidth of SZ-1.4 compared with the pSZ baseline as well as the best performing configuration of vecSZ. For vecSZ, we use the best configuration of blocksize and vector length, found in Section~\ref{sec:vec_perf}.  We break down the performance on the AMD and Intel CPUs in Figure~\ref{fig:version_comparison_amd} and Figure~\ref{fig:version_comparison_intel}, respectively.

\begin{figure}[h!]
\begin{subfigure}{.49\linewidth}
  \centering
  \includegraphics[width=\linewidth, trim={16 16 16 15}, clip]{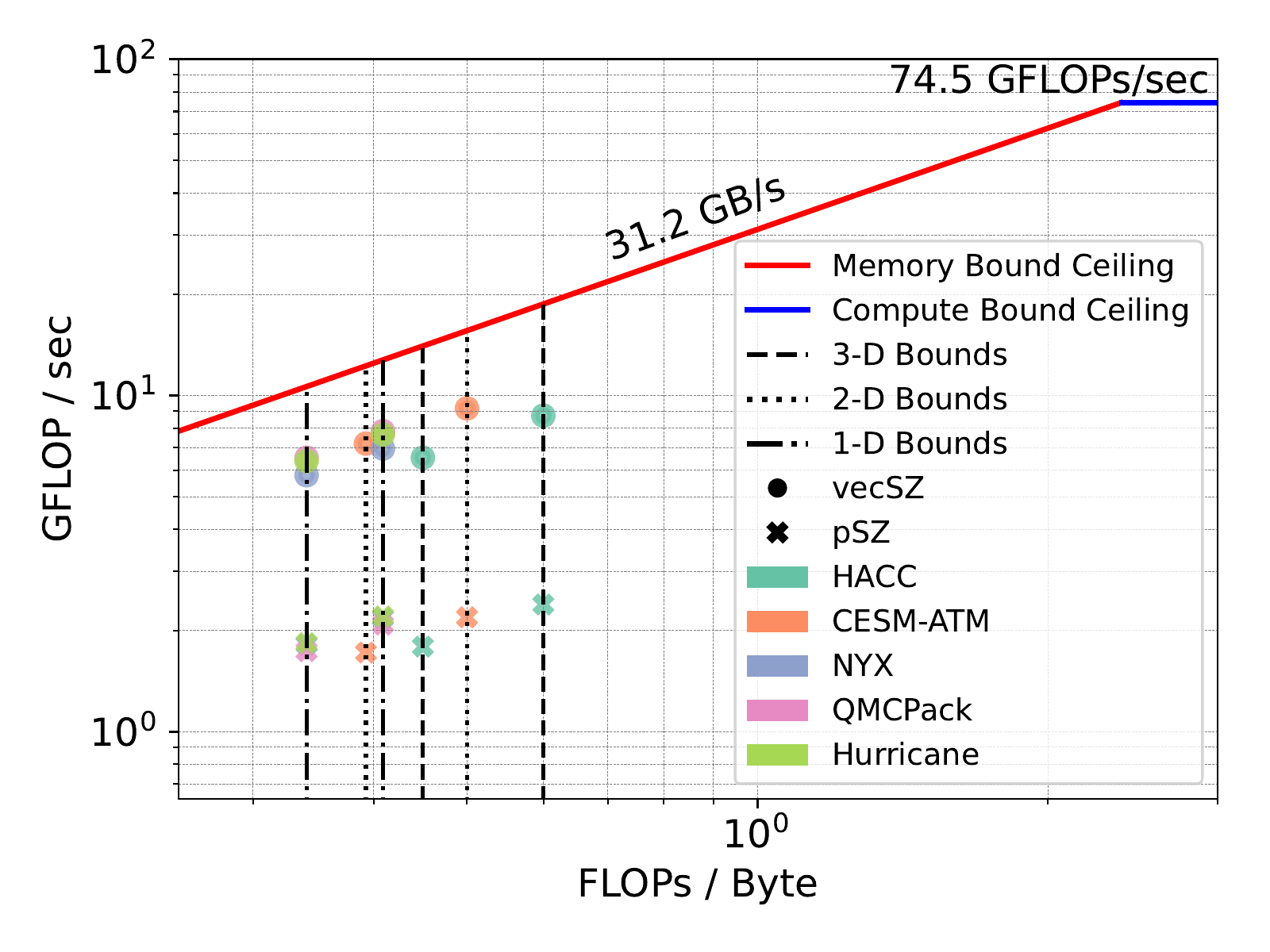}  
  \caption{AMD}
  \label{fig:roofline-amd}
\end{subfigure}
\begin{subfigure}{.49\linewidth}
  \centering
  \includegraphics[width=\linewidth, trim={16 16 16 15}, clip]{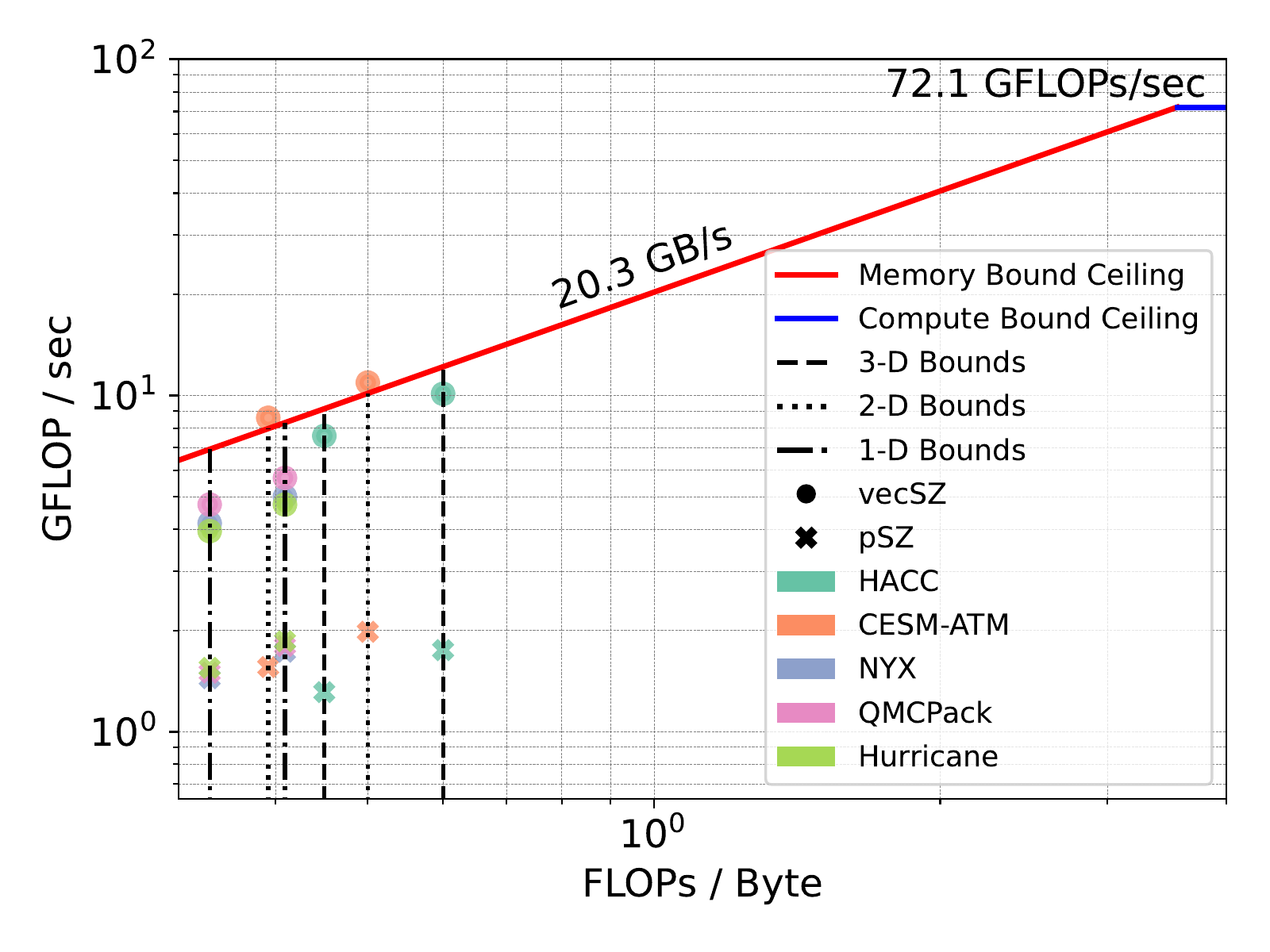}  
  \caption{Intel}
  \label{fig:roofline-intel}
\end{subfigure}
\caption{Roofline Model showing vecSZ performance for the Dual-Quantization algorithm for vecSZ compiled with O3 and no vectorization baseline (pSZ) and vecSZ with compiled with O3 and vectorization enabled. }
\label{fig:roofline}
\end{figure}

For the AMD CPU, we see that the pSZ baseline shows better performance, on average, than SZ-1.4 in three of the five applications, with a maximal speedup of 3.4$\times$ for HACC. The AMD 7452 shows its best performance for HACC and CESM-ATM, improving over the pSZ baseline by at least 2000 MB/s, resulting in speedup of 3.7$\times$ and 4.1$\times$ respectively. Compared to SZ-1.4, vecSZ improves prediction and quantization bandwidth by 2700 MB/s for 1D and 2200 MB/s for 2D data, for a speedup of 13.1$\times$ and 12$\times$ for 1D and 2D data respectively.  For 3D data sets, we outperform SZ-1.4 by between 2.2$\times$ and 7.3$\times$ and pSZ by between 3.2$\times$ and 3.7$\times$ depending on the data set. This corresponds to increases in prediction and quantization bandwidth of between 1000 MB/s and 1100 MB/s for SZ-1.4 and 900 MB/s and 1200 MB/s for pSZ.
3D data sets result in increased prediction and quantization bandwidth on the AMD CPU as opposed to the Intel CPU due to it's larger cache, resulting in an average of 30\% fewer cache misses for higher dimensional data sets. 

For the Intel CPU, we see that pSZ has better performance than SZ-1.4 in the same three applications. Furthermore, we find that for the 1D and 2D test data sets, our vectorized code increases prediction and quantization bandwidth by greater than  2800 MB/s for SZ-1.4 and 2600 MB/s for pSZ, resulting in a 15.1$\times$ speedup over SZ-1.4 and 6$\times$ speedup over pSZ for 1D data sets, and 14.0$\times$ and 5.7$\times$ for SZ-1.4 and pSZ, respectively, on 2D data sets. Three-dimensional data sets exhibit less performance improvement compared to 1D and 2D data sets, resulting in an increase of between 600 and 900 MB/s or speedup between 2.3$\times$ and 5.3$\times$ depending on data set. The decrease in speedup when performing dual-quant for 3D data is due to the organizational pattern of 3D data as opposed to that of 1D data which has the potential to result in up to 6$\times$ the number of cache misses.

Both the AMD and Intel CPUs display similar trends in their prediction and quantization bandwidth across data sets of different dimensions. Comparing the performance of the CPUs, we see that for 2D data sets the larger vector registers of the Intel Gold enable it to outperform the AMD 7452. However, the larger cache of the AMD 7452 leads to higher prediction and quantization bandwidth for 3D data sets than the Intel CPU provides. Overall, we find that vecSZ improves upon the prediction and quantization bandwidth of SZ-1.4 by 8.7$\times$ for AMD and 9.2$\times$ for Intel on average, with a peak prediction and quantization bandwidth in excess of 2.9 GB/s for both AMD and Intel CPUs.

\subsubsection{Roofline Analysis}
As introduced in Section \ref{sec:roofline}, to quantify how much of peak performance our vectorized code obtains we compare our results to the expected performance from the Roofline model.
Using the peak GFLOPs attained by the optimal configuration of block size and vector length, we plot the dual-quant performance of pSZ and our vectorized SZ using the roofline model (see Section \ref{sec:roofline}). 

The AMD 7452 CPU's performance results in peak percentage of DRAM memory bandwidth between 47-61\%, improving over the baseline O3 optimized code by 3.2--4.2$\times$. The Intel Gold CPU shows between 57-107\% of peak DRAM memory bandwidth. HACC and CESM-ATM result in the closest to peak performance. CESM-ATM shows a GFLOP/sec value over the peak DRAM memory bandwidth because it is able to fit within the 22 MB L3 cache, whereas larger data sets cannot. We do not see this behavior for CESM on the AMD CPU because it is able to sustain a higher peak memory bandwidth and does not have support for 512-bit vector registers that provide the best performance on the Intel CPU. The combined effect of these two features results in the performance difference between Figures \ref{fig:roofline-amd} and \ref{fig:roofline-intel}. For all applications and both CPUs, we consistently improve over the baseline, pSZ, by 2.5--5.6$\times$, depending on the data set.

\subsection{Understanding Vectorized Performance} \label{sec:vec_perf}
The performance of the dual-quant operation is largely dependent on the block size used for chunking input data sets, as well as the length of the vector registers used in computation. To determine the optimal configuration of block size and vector length for each data set, we perform an exhaustive run of all possible block size and vector length configurations. We show the average prediction and quantization bandwidth per application for each of the possible configurations in Figure~\ref{fig:blksz}.



\subsubsection{Block Size and Vector Length}
\begin{figure}[h!]
\begin{subfigure}{.49\linewidth}
  \centering
   \includegraphics[width=\linewidth, trim={14 15 13 14}, clip]{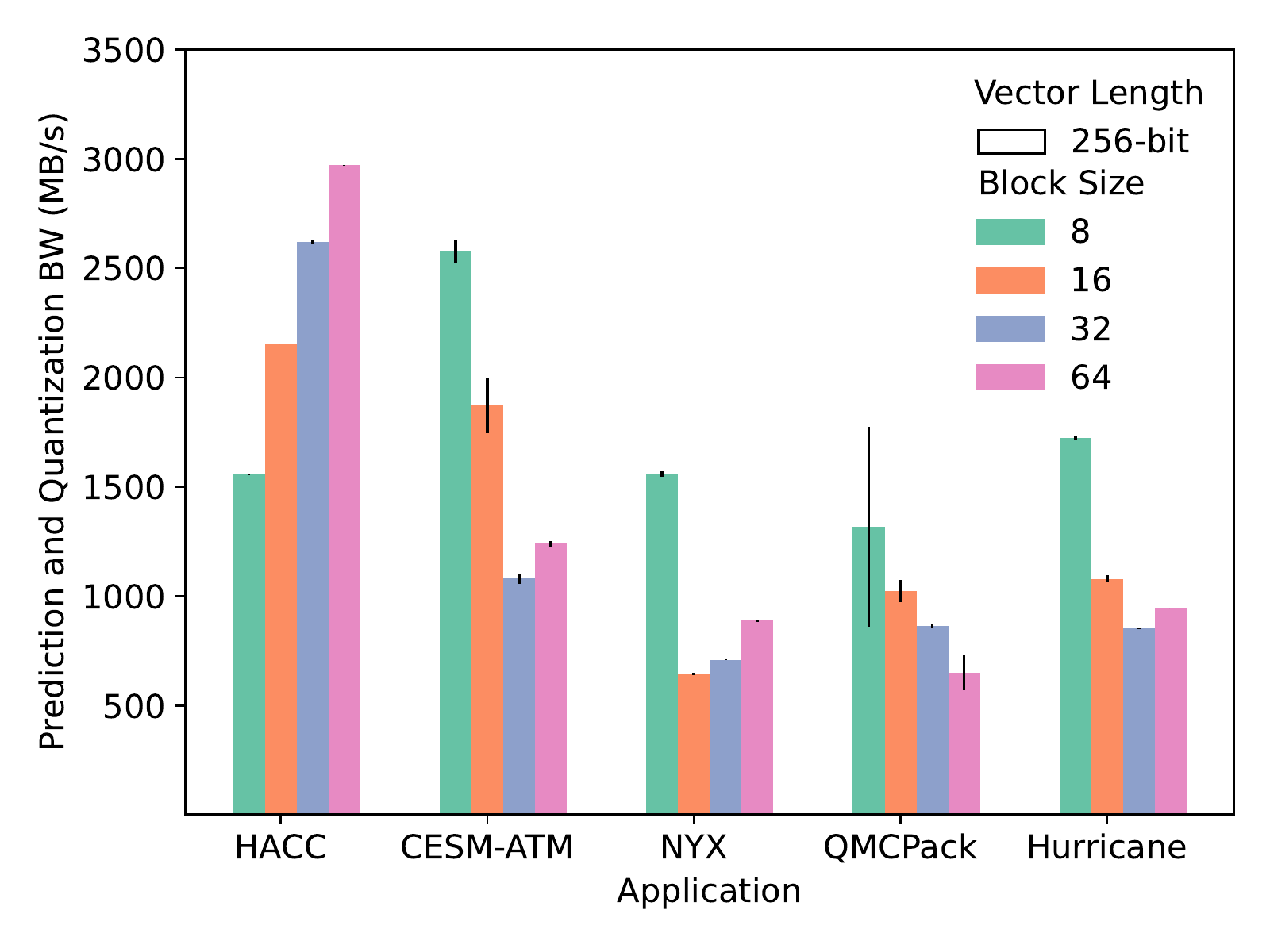}
  \caption{AMD}
  \label{fig:blksz-amd}
\end{subfigure}
\begin{subfigure}{.49\linewidth}
  \centering
   \includegraphics[width=\linewidth, trim={14 15 13 14}, clip]{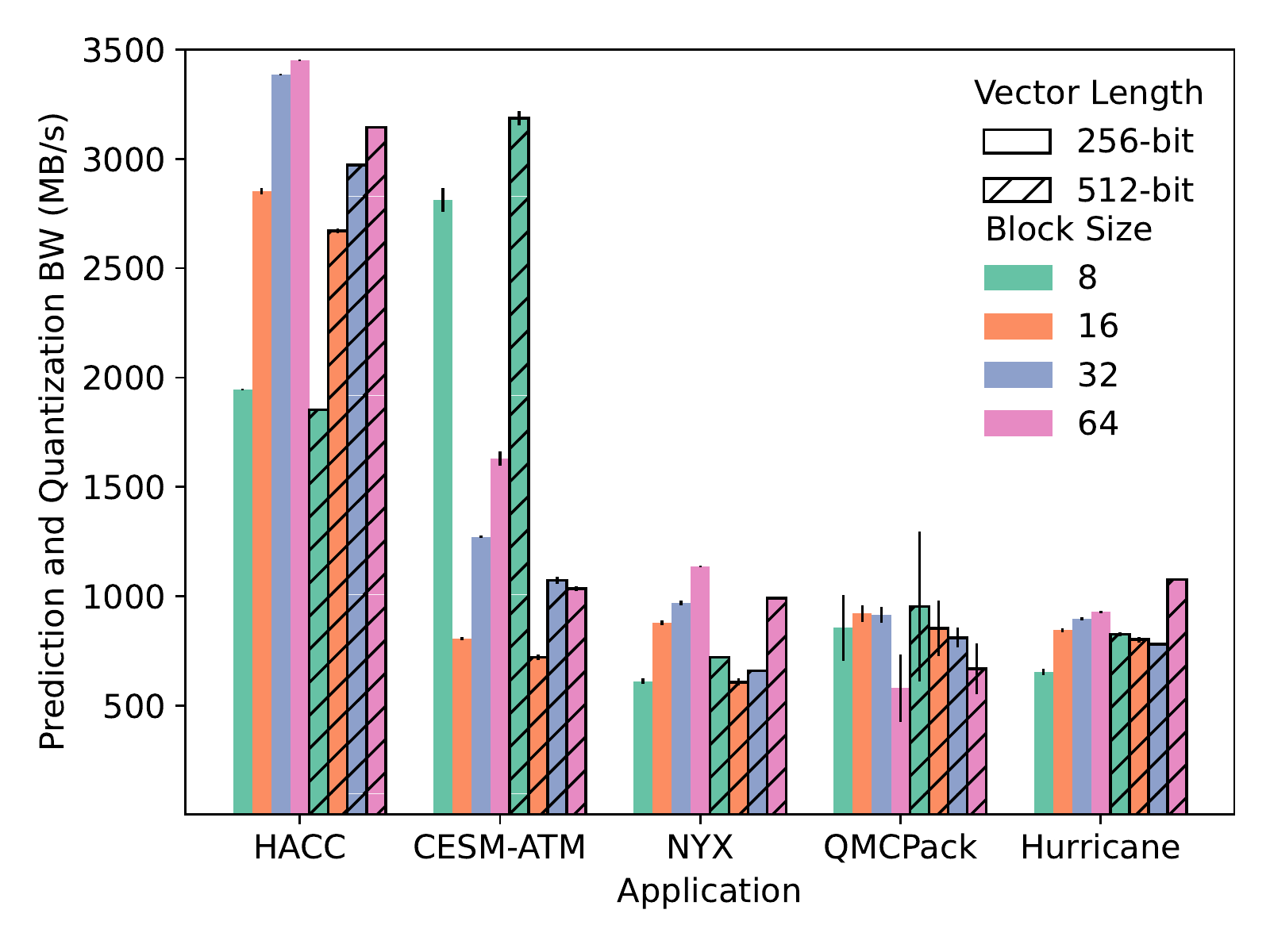}
  \caption{Intel}
  \label{fig:blksz-intel}
\end{subfigure}
\caption{Performance for different vector register lengths and compression block sizes.}
\label{fig:blksz}
\end{figure}

Block size has the largest impact on performance of the dual-quant operation. In some cases, adjusting the block size improves the prediction and quantization bandwidth by 354\% for the Intel and 139\% for the AMD CPUs. The differences in block size performance also vary across applications. For example, the HACC experiments continue to improve as block size increases, whereas the general trend of QMCPack tends to decrease as block size increases. 
For AMD, we find that a block size of 64 performs best for 1D data sets while a block size of 8 leads to optimal performance for all other 2D and 3D data sets. The Intel performance shown in Figure \ref{fig:blksz-intel} presents a more ambiguous result which varies depending on vector length and the data set on which dual-quant is being performed. To account for this variability we explore autotuning in Section~\ref{sec:autotune}. 

Since the AMD 7452 only supports up to 256-bit vector registers, the length of the maximum vector registers we use is a configuration option only for the Intel CPU. In most cases, vector registers of size 256 perform slightly better for the Intel CPU. However, for CESM, QMCPack, and Hurricane, 512-bit vector registers is best. A potential reason for this improved performance of 256-bit over 512-bit vector registers is that the size of a cache line in the Intel CPU is 64 bytes. This means that one 512-bit vector contains an entire cache line worth of data. If data is misaligned, it results in degraded performance.

Block size must be a multiple of the vector register in use. With respect to the performance of block sizes of 8 and vector lengths of 512-bits for the Intel CPU, a block size of 8 is not a multiple of the vector register length. In this case, autovectorization performed by the compiler uses 512-bit vector registers while our manual vectorization reverts to a vector size of 256-bits in order to utilize the entire vector register. This creates a hybrid of 512-bit and 256-bit vector operations that attribute to the performance improvements between 512-bit and 256-bit vector lengths for block size 8.

\subsection{Autotuning Block Size and Vector Length} \label{sec:autotune}
Section~\ref{sec:vec_perf} shows that the performance of vecSZ depends on the configuration of block size and vector width. Intel CPUs have a total of 8 potential configurations of block size and vector length, while AMD CPUs have a total of 4 configurations of block size. We apply an autotuning algorithm to vecSZ to determine the best performing block size and vector length based on the best average performance of the dual-quant operation on a subset of all blocks in a data set, completed across multiple iterations. 

\begin{figure}[h!]
\begin{subfigure}{.49\linewidth}
  \centering
  \includegraphics[width=\linewidth, trim={15 17 20 14}, clip]{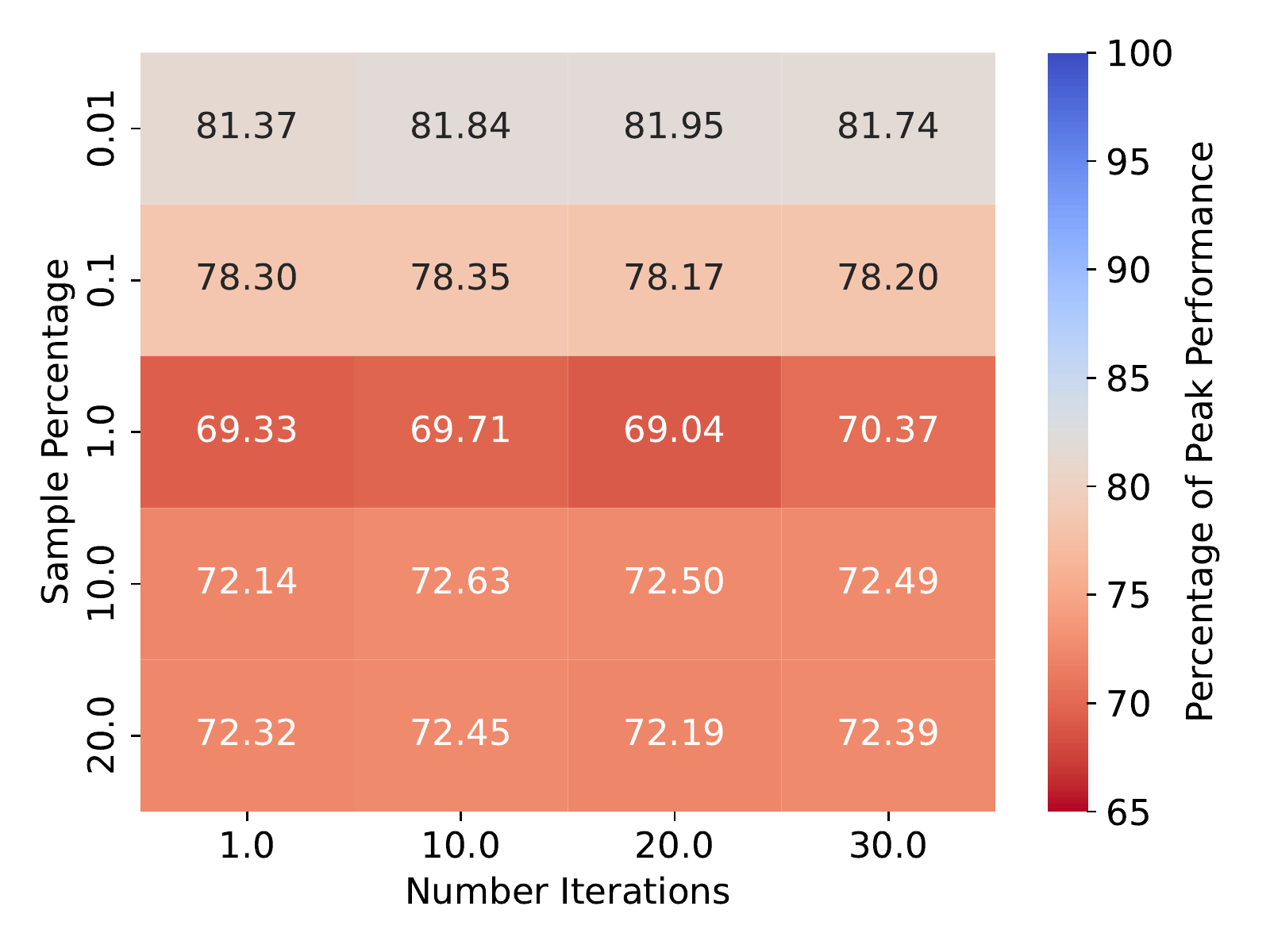}  
  \caption{AMD}
  \label{fig:autotune-peak-amd}
\end{subfigure}
\begin{subfigure}{.49\linewidth}
  \centering
  \includegraphics[width=\linewidth, trim={15 17 20 14}, clip]{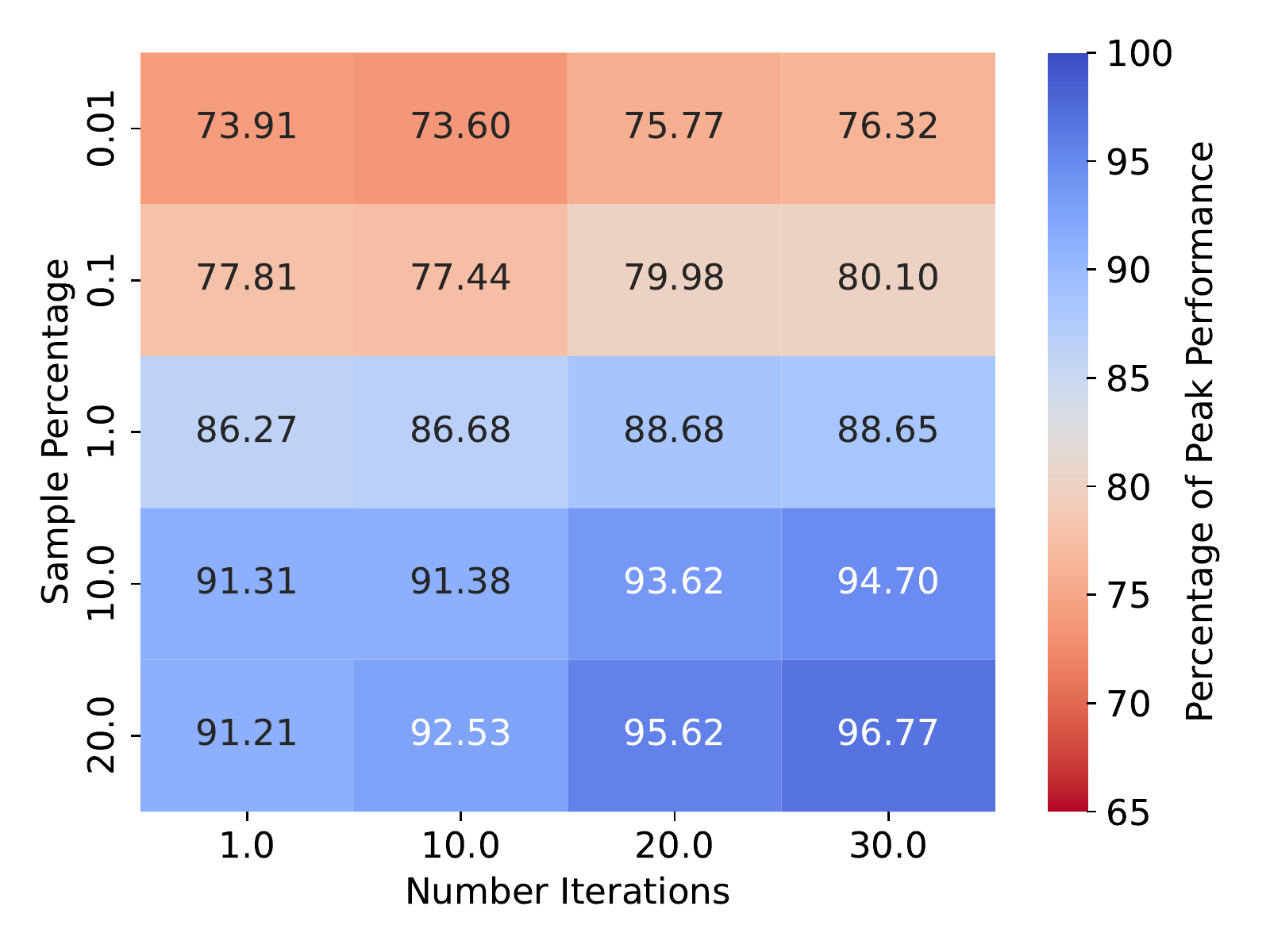}  
  \caption{Intel}
  \label{fig:autotune-peak-intel}
\end{subfigure}
\caption{Percentage of peak performance achieved through autotuning.}
\label{fig:autotune-perf}
\end{figure}

Figure \ref{fig:autotune-perf} states the percentage of the peak performance of a configuration achieved by each pair of autotuning settings. In these figures, the more frequently an autotuning run results in selection of the correct configuration of block size and vector length, the closer that configuration's portion of the heatmap is to the peak performance. The values presented in these figures are averaged across all applications Overall, results for each individual application follows a comparable trend in percentage of peak prediction and quantization performance. Figure \ref{fig:autotune-peak-amd} results in a higher percentage of peak performance for smaller sample percentages of blocks for the AMD CPU. Since across iterations the same blocks are being computed, smaller sample percentages are more likely to already have all their blocks resident in cache, although a random sample of data is begin taken. Since the cache of the Intel Gold CPU is much smaller, this issue is not present. This allows a more accurate configuration to be found via autotuning as more samples are taken and averaged over a larger number of trials.

\begin{figure}[h!]
\begin{subfigure}{.49\linewidth}
  \centering
  \includegraphics[width=\linewidth, trim={15 17 20 14}, clip]{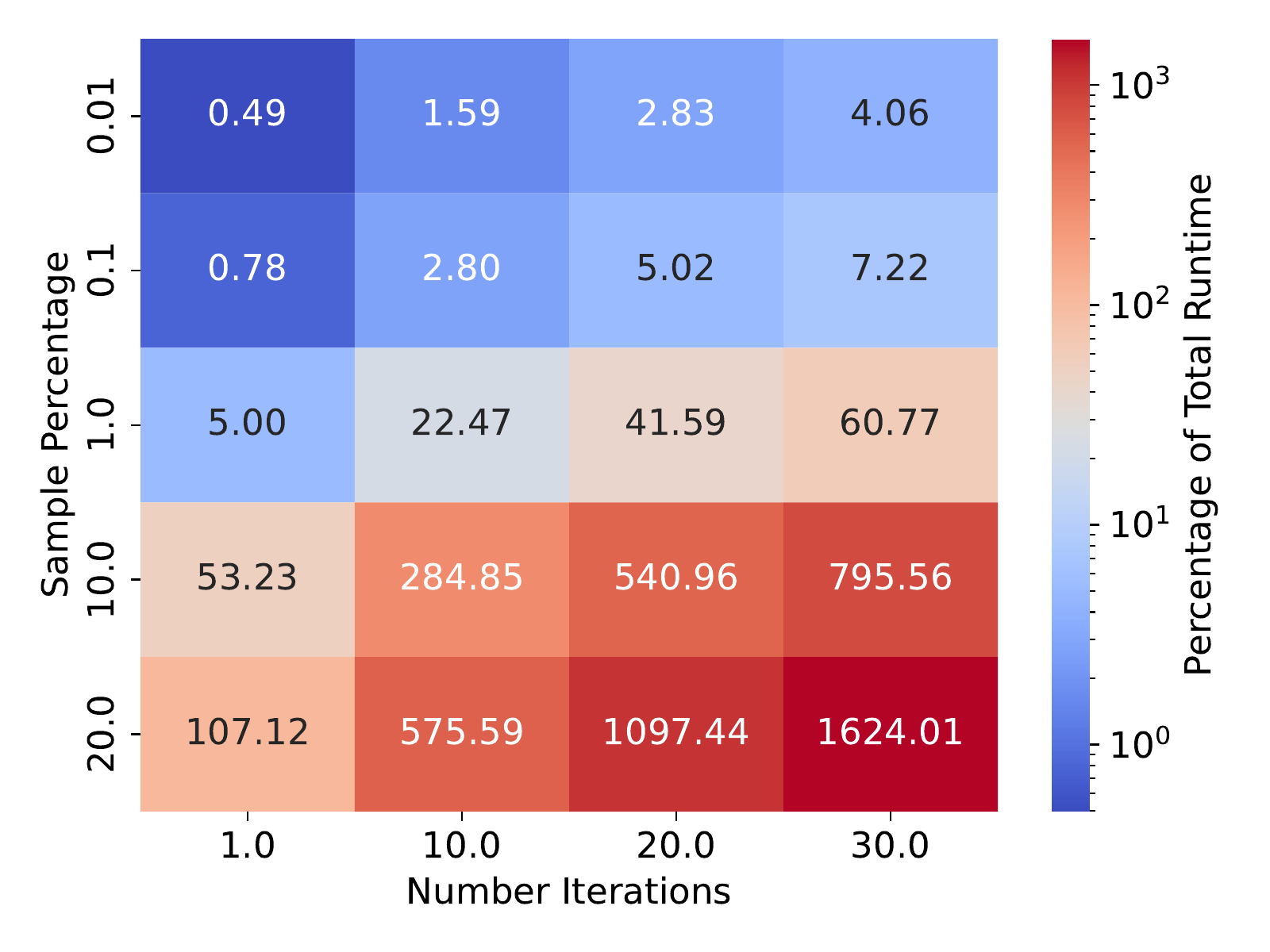}  
  \caption{AMD}
  \label{fig:autotune-time-amd}
\end{subfigure}
\begin{subfigure}{.49\linewidth}
  \centering
  \includegraphics[width=\linewidth, trim={15 17 20 14}, clip]{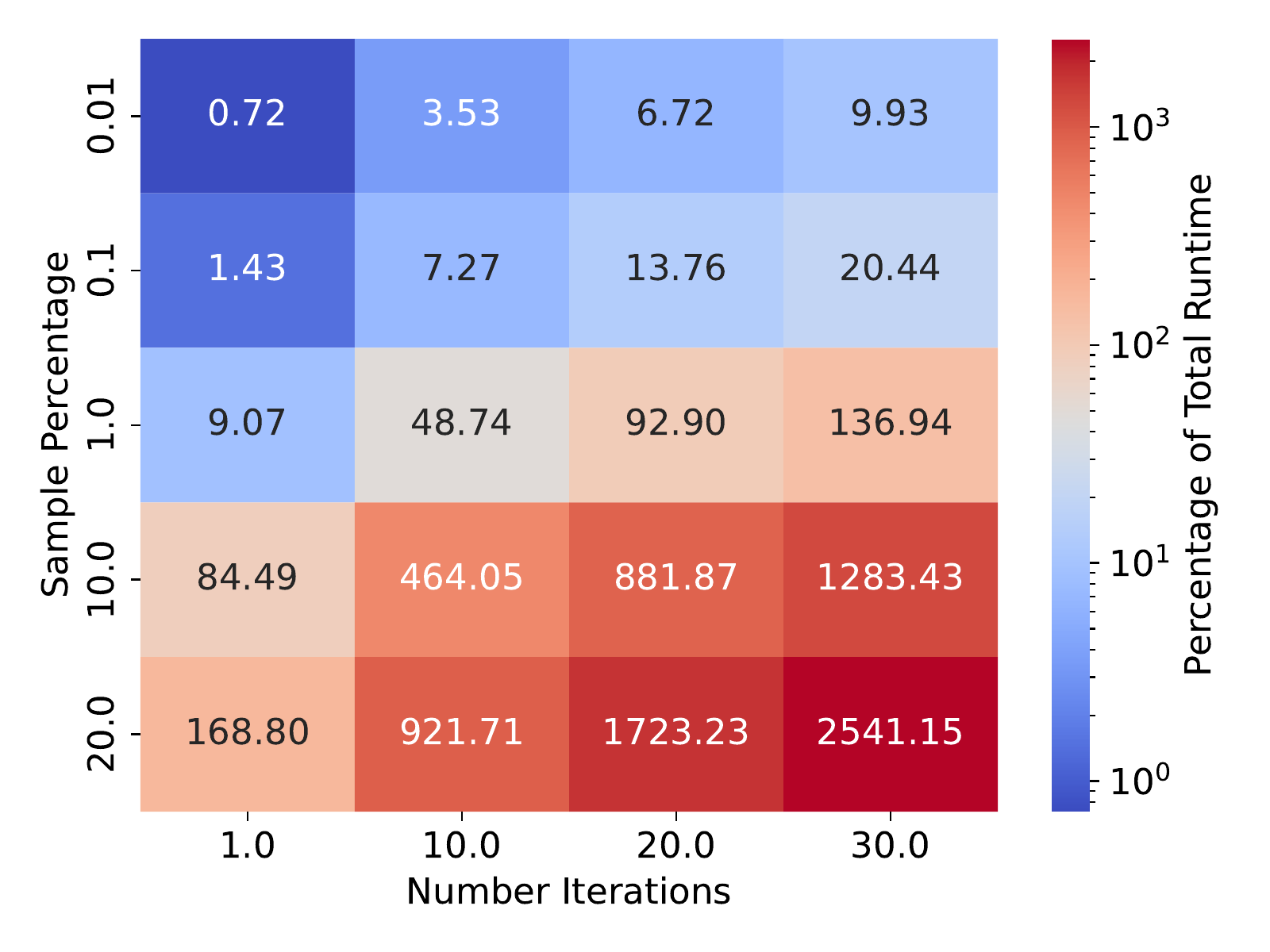}  
  \caption{Intel}
  \label{fig:autotune-time-intel}
\end{subfigure}
\caption{Percentage of optimal runtime spent on autotuning.}
\label{fig:autotune-time}
\end{figure}

Autotuning results for both CPUs show that, as the percentage of blocks sampled and the number of iterations increases, the percentage of total runtime spent autotuning increases as well. The AMD 7452 requires a lower percentage of the total runtime on autotuning for all autotuning configurations because of the reduced number of configurations that must be tuned since 512-bit vector lengths are not supported. For the Intel CPU, we compare results for each block size and vector register length, which results in a larger percentage of total runtime spent on autotuning. In Figure \ref{fig:autotune-time}, we present that increasing the number of iterations performed before averaging results in less time cost than increasing the percentage of the blocks sampled, except when performing one iteration and moving from a sample percentage of 10\% to 20\%. 

When choosing a reasonable percentage of blocks to sample and iterations to average before autotuning, it is also important to consider the time required to  autotune for a given configuration. Depending on the application, a trade-off to consider is a balance between gaining closer to peak performance and spending less time tuning the configuration. 
Figure \ref{fig:autotune-time-amd} and \ref{fig:autotune-time-intel} show the percentage of the total runtime spent autotuning depending on the percentage of all blocks that is sampled and the number of iterations run before taking the average of dual-quant times to find the best configuration.

The Intel CPU exhibits a trend corresponding with the result that if a larger percentage of blocks from a data set are sampled, the tuning configuration chosen by our autotuning algorithm is closer to optimal. Additionally, if we repeat our experiment for a larger number of iterations, we achieve a value within 5--10\% of peak performance. The high-cost of autotuning for larger percentages and larger number of iterations could be amortized over several compressor configurations as recent work has shows that for time-series data~---~e.g., fields in a scientific application~---~a optimal compressor configuration remains mostly constant through time~\cite{underwood2020fraz}. We find, for example, that across all 48 time-steps of a field of the Hurricane Isabel dataset, an average of 80\% of the autotuning runs result in two top configurations. Using this knowledge, we can autotune based on the top two configurations to drastically reduce the overhead by narrowing the possible configurations.

\subsection{Thread-level Parallelization}
\begin{figure}[h!]
\begin{subfigure}{.49\linewidth}
  \centering
   \includegraphics[width=\linewidth,trim={1cm 0 1cm 0}]{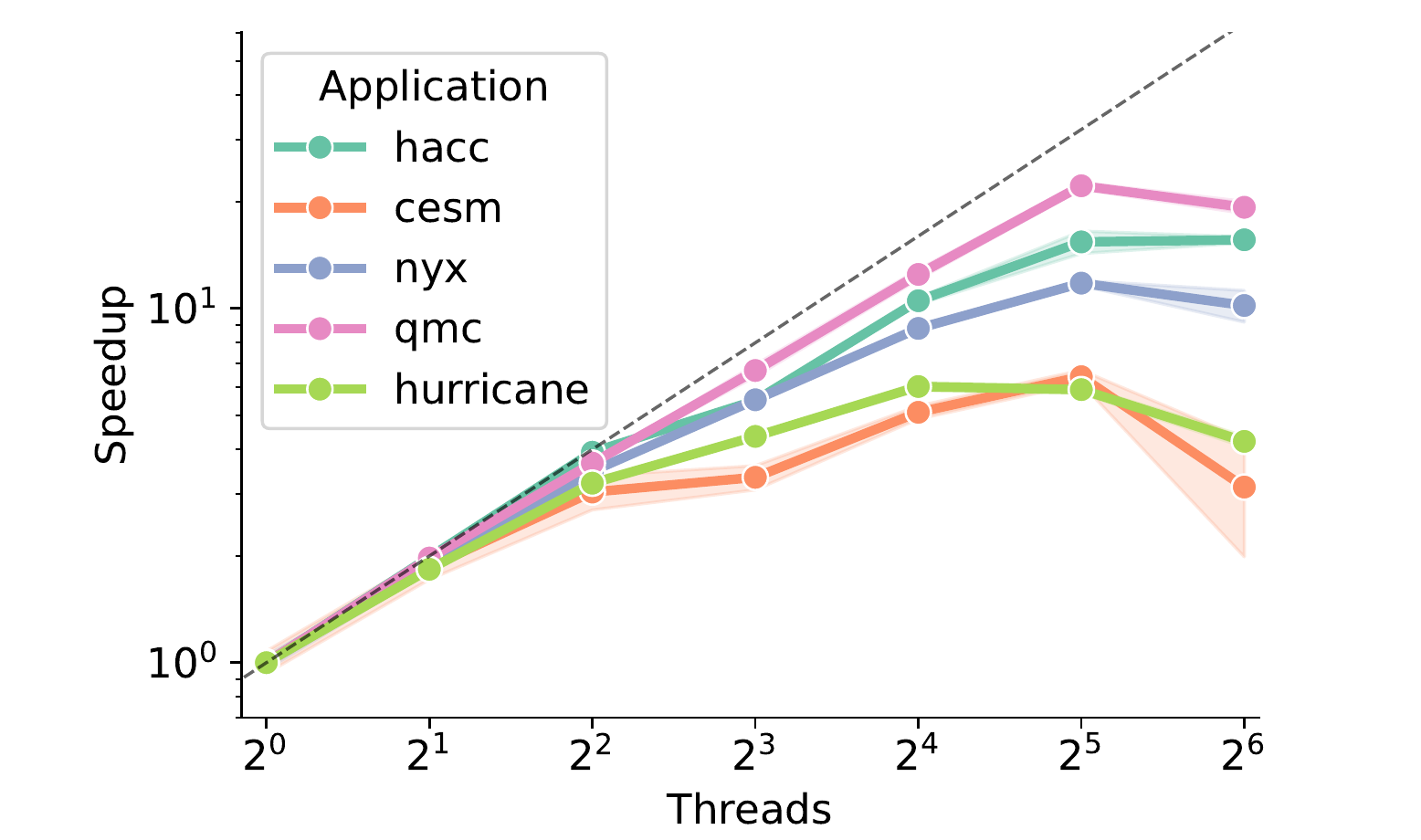}  
  \caption{AMD}
  \label{fig:omp-amd}
\end{subfigure}
\begin{subfigure}{.49\linewidth}
  \centering
  \includegraphics[width=\linewidth,trim={1cm 0 1cm 0}]{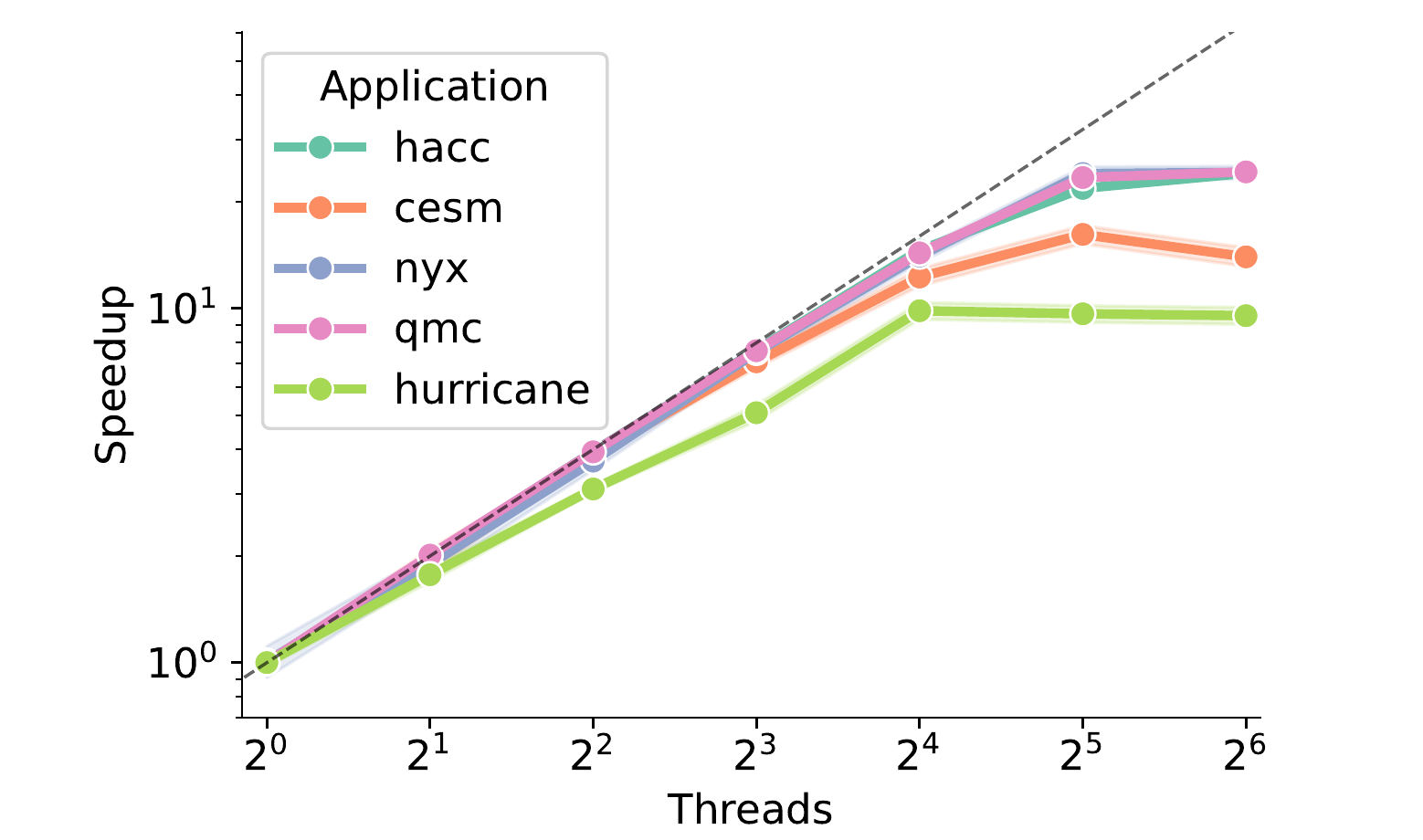}  
  \caption{Intel}
  \label{fig:omp-intel}
\end{subfigure}
\caption{Single node OpenMP scaling performance.}
\label{fig:omp}
\end{figure}

We perform thread-level parallelization at the block level for the dual-quant operation in vecSZ using OpenMP. We perform scaling tests using between 1 and 64 threads for each CPU, and present the speedup of vecSZ for each application over the single-threaded performance of vecSZ in Figure \ref{fig:omp}. Overall, we see a maximal speedup of 24$\times$ at 64 threads. 

For the AMD 7452, we find that for up to 4 threads, the dual-quant operation scales nearly linearly. Once we reach 8 OpenMP threads, CESM-ATM and Hurricane data sets begin to reach their peak speedup, while HACC, NYX, and QMCPack continue scaling linearly until 64 threads where they begin to plateau. The down tick in performance between 32 and 64 threads is attributed to the number of cores. As we explain in section \ref{sec:hardware}, the AMD CPU we perform scaling tests on has 32 cores in total, with the ability to run up to 64 threads. When moving from using 32 threads to 64, each CPU core adjusts to run two threads, resulting in a performance decrease from the previous number of threads tested.

For all applications with the exception of Hurricane, the Intel CPU scales close to linearly until reaching 16 threads. Hurricane's begins to not scale well at 4 threads, but at 16 we see performance plateaus. When moving from 16 to 32 OpenMP threads, we begin to utilize the second CPU on the node, resulting in slight deviations from the previous trends of the data set. When operating at 64 threads, we see down ticks similar to those observed in Figure \ref{fig:omp-amd}, where performance is affected by the presence of multiple threads per core. However, we see that as we leverage the hyper-threading, the scaling performance does not deteriorate to the same degree as using the simultaneous multi-threading on the AMD CPU.

\begin{figure}
\begin{subfigure}{.49\linewidth}
  \centering
  \includegraphics[width=\linewidth, trim={18 18 19 10}, clip]{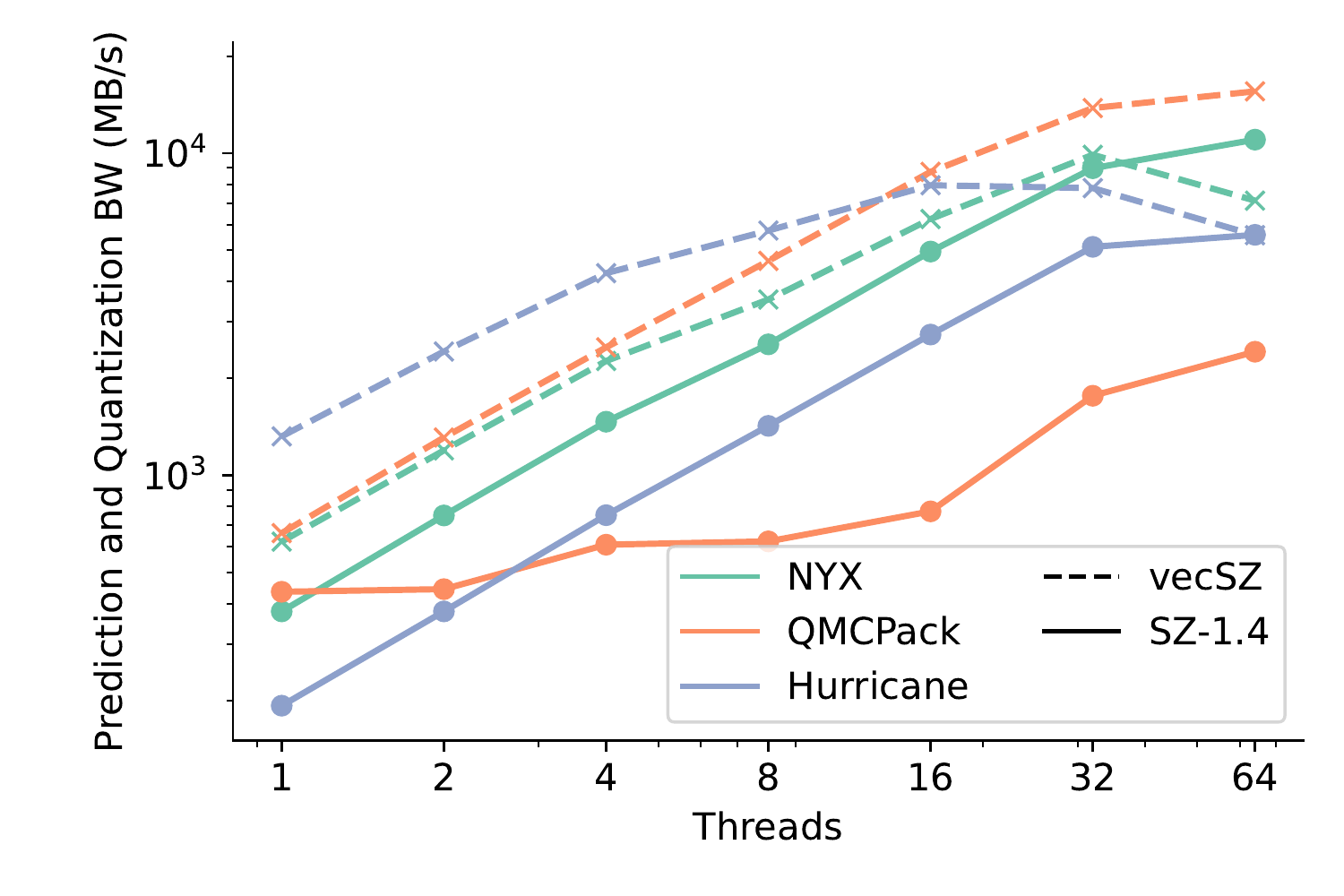}  
  \caption{AMD}
  \label{fig:omp-sz-amd}
\end{subfigure}
\begin{subfigure}{.49\linewidth}
  \centering
  \includegraphics[width=\linewidth, trim={15 18 19 10}, clip]{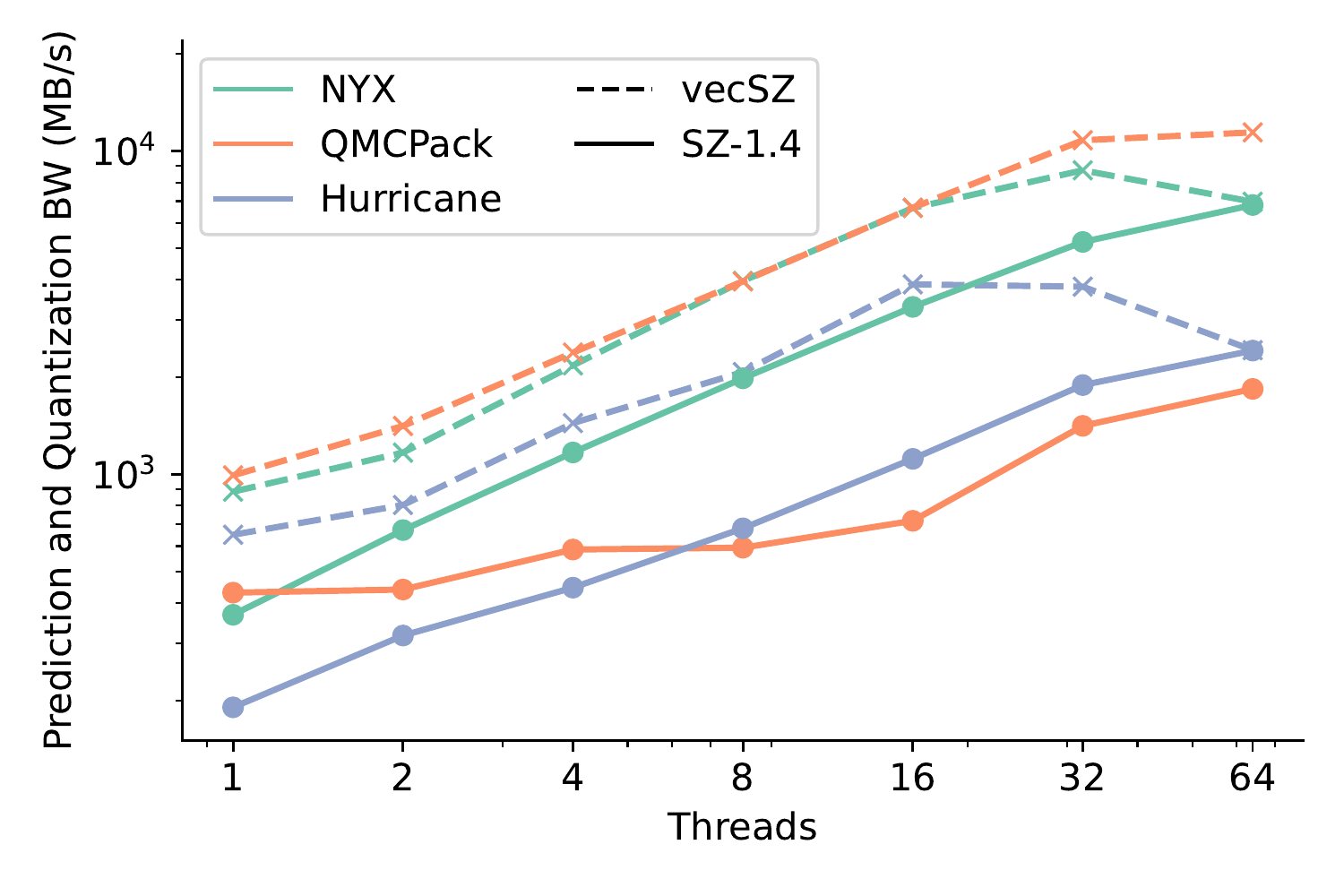}  
  \caption{Intel}
  \label{fig:omp-sz-intel}
\end{subfigure}
    \caption{OpenMP performance for vecSZ and SZ-1.4}
    \label{fig:sz14_omp}
\end{figure}

\subsubsection{Comparison to SZ-1.4 OpenMP Scaling}
Additionally, to explore how vecSZ performs with respect to SZ-1.4, we compare of vecSZ's prediction and quantization bandwidth to that of SZ-1.4, shown in Figure \ref{fig:sz14_omp}. We only show the performance for the 3D data sets as SZ-1.4 does not support OpenMP compression of 1D or 2D data sets.

When performing the prediction and quantization operation on up to 32 threads, vecSZ outperforms SZ-1.4 by as much as 11.6$\times$ for 16 threads for QMCPack on the AMD CPU. For some data sets; however, SZ-1.4 outperforms vecSZ when running with 64 OpenMP threads. The degraded prediction and quantization bandwidth of vecSZ when scaling with a higher number of threads is due to the amount of resources taken up by vectorization. Using 64 threads decreases the amount of data each thread operates on. As the amount of work per thread decreases, the likelihood of false sharing and contention in private caches increases. This leads to a decrease in performance as vecSZ needs more data per instruction than SZ-1.4, further complicating these issues.

\subsection{Overall Impact}

In the results presented above, we focus on optimizations performed on the prediction and quantization operations of the compression pipeline. We now address the impact of our optimizations on the overall performance of vecSZ. 

\begin{table}[ht]
    \centering
    \begin{tabular}{ccc}
        & \multicolumn{2}{c}{\textbf{CPU}}\\
        & \textit{AMD EPYC Rome} & \textit{Intel Xeon Gold} \\
        \toprule
        \multicolumn{1}{c}{\textbf{Dual-Quant \% of Runtime}} & 46.9\% & 42.9\%\\
        \multicolumn{1}{c}{\textbf{Theoretical Max Speedup}} & 1.70$\times$ & 1.67$\times$\\
        \multicolumn{1}{c}{\textbf{Actual Speedup}} & 1.51$\times$ & 1.47$\times$\\
        \multicolumn{1}{c}{\textbf{\% of Theoretical Achieved}} & 88.9\% & 87.6\%\\
        \bottomrule
    \end{tabular}
    \caption{Theoretical and actual speedup for vecSZ over it's serial implementation, pSZ, averaged for all applications.} 
    \label{tab:overall_speedup}
\end{table}

The dual-quant operation takes an average of 46.9\% and 42.9\% of the total sequential runtime for the AMD and Intel CPUs respectively. Using Amdahl's Law, 
theoretical maximum speedup is computed by: $S = \frac{1}{(1-p) + p/s}$, where
$p$ is the proportion of the sequential runtime that an operation takes and $s$ is the speedup of the operation being parallelized. For the AMD CPU, we set $s=8$ because a 256-bit vector register fits 8 32-bit floating point values. Because the Intel CPU has 512-bit vector registers, we set $s=16$.

Using this equation, we find the theoretical speedup shown in Table \ref{tab:overall_speedup} of the total runtime possible by performing vectorization of the dual-quant operation to be 1.7$\times$ for AMD and 1.67$\times$ for Intel. We achieve 89\% of the theoretical maximum by reaching a total speedup of 1.51$\times$ for AMD and 88\% of our theoretical maximum speedup at 1.47$\times$ for Intel.

\begin{figure}
    \centering
    
    \includegraphics[width=.75\linewidth, trim={15 15 20 50}, clip]{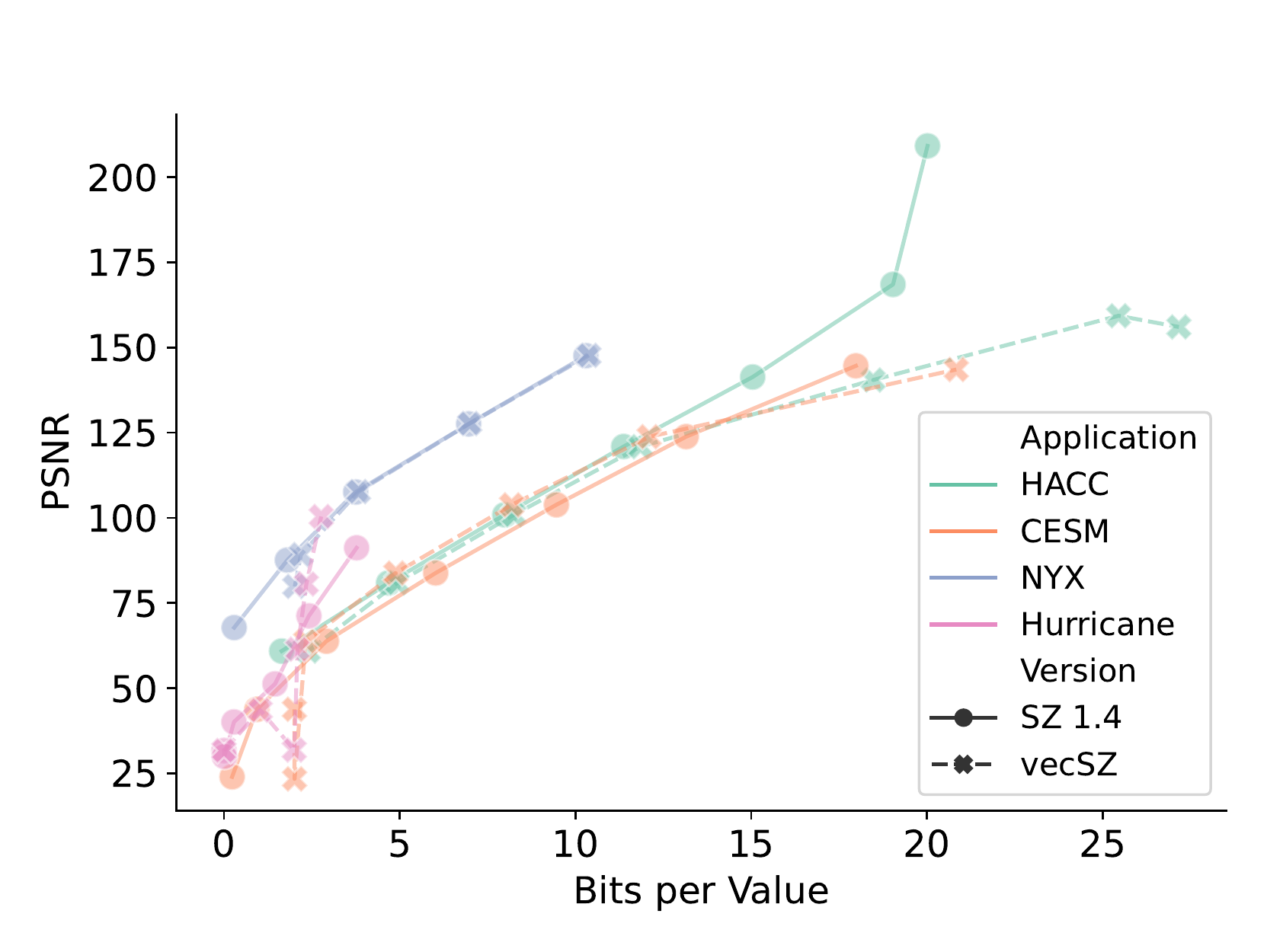}
    \caption{Rate distortion comparison of vecSZ and SZ-1.4.}
    \label{fig:rate distortion}
\end{figure}

\subsection{Impact of Alternative Padding}
Across all block sizes and error-bounds, the use of alternative padding values results in decrease in unpredictable values by an average of 10\%. The best cases results in 100\% elimination of all outlier data. For larger error-bounds, the prediction of border values becomes more important because fewer unpredictable values exist in the rest of the compression block, therefore a larger percentage of the unpredictable values come from border regions that remain unpredictable regardless of error-bound due to the use of zero padding. In these cases, application of a global or block average as alternative padding results in elimination of all outliers because we make it possible to predict the border values. We also find that use of minimum and maximum values for padding do not perform as well as average values. These do not serve as good metrics for determining an alternative padding value because they commonly tend to be outliers in the dataset, thus they tend to poorly represent the values along the border. 

Using a global average value for padding to provide padding that is representative of the data without incurring excess overhead of storing additional padding values we generate Figure~\ref{fig:rate distortion}, which shows the rate-distortion of vecSZ and SZ-1.4. For reasonable error-bounds, we perform equally, or better than SZ-1.4. For CESM and Hurricane data, we improve rate-distortion up to 18.9\% and 32\% respectively. This improvement can be further advanced by future work, such as implementation of an improved compression method for remaining outlier data, which we currently handle via a lossless compressor pass.



\section{Conclusions and Future Work} \label{sec:conclusion}

As large-scale applications generate larger volumes of data, compression techniques are needed to ensure the data is able to be stored and transmitted effectively. In this paper, we present and optimize vecSZ, a threaded and vectorized version of the dual-quantization GPU algorithm for CPU architectures. We find that best performance is dependent on data set, compression block size, and vector register length.  The autotuning of vecSZ selects the optimal block size and vector width ensuring best performance. 

Results show that vecSZ improves the prediction and quantization bandwidth by up to 15.1$\times$ compared to SZ-1.4, improving peak DRAM bandwidth by as much as 61--107\%. Our novel non-zero block padding reduces the number of outlier data on the block's border by up to 100\% yielding up to a 32\% improvement in rate distortion. 
Future work will explore how vecSZ runs on non x86 architectures such as ARM architectures that leverage scalable vector extensions.



\bibliographystyle{IEEEtran}
\bibliography{refs}

\end{document}